\documentclass[aps,pra,showpacs,onecolumn, amsfonts,amssymb,amsmath]{revtex4}  
\usepackage{graphicx} 
\usepackage{array}
\usepackage{color}
\usepackage{float}
\usepackage{subfigure}
\usepackage{color}
\usepackage{multirow}

\newcommand{\YSO}[0]{Y$_2$SiO$_5\,$}
\begin{document}

\title{Fast and robust creation of an arbitrary single qubit state by nonadiabatic shortcut pulses in a three-level system} 

\author{Ying Yan$^1$\footnotemark[1], Yi Chao Li$^2$, , Adam Kinos$^3$, Andreas Walther$^3$, Chunyan Shi$^3$, Lars Rippe$^3$, Joel Moser$^1$, Stefan Kr\"{o}ll$^3$, Xi Chen$^1$\footnotemark[1]} %
\affiliation{$^1$ College of Physics Optoelectronics and Energy and Collaborative Innovation Center of Suzhou Nano Science and Technology; Key Lab of Advanced Optical Manufacturing Technologies of Jiangsu Province and Key Lab of Modern Optical Technologies of Education Ministry of China, Soochow University, 215006 Suzhou, China}
\affiliation{$^2$ Department of Physics, Shanghai University, 200444 Shanghai, China}
\affiliation{$^3$ Department of Physics, Lund Institute of Technology, P.O.~Box 118, SE-22100 Lund, Sweden}
%
\footnotetext[1]{yingyan@suda.edu.cn, xchen@shu.edu.cn}

\date{\today}

\begin{abstract}
High-fidelity qubit initialization is of significance for efficient error correction in fault tolerant quantum algorithms. Combining two best worlds, speed and robustness, to achieve high-fidelity state preparation and manipulation is challenging in quantum systems, where qubits are closely spaced in frequency. Motivated by the concept of shortcut to adiabaticity, we theoretically propose the shortcut pulses via inverse engineering and further optimize the pulses with respect to systematic errors in frequency detuning and Rabi frequency. Such protocol, relevant to  frequency selectivity, is applied to rare-earth ions qubit system, where the excitation of frequency-neighboring qubits should be prevented as well. Furthermore, comparison with adiabatic complex hyperbolic secant pulses shows that these dedicated initialization pulses can reduce the time that ions spend in the excited state by a factor of 6, which is important in coherence time limited systems to approach an error rate manageable by quantum error correction. The approach may also be applicable to superconducting qubits, and any other systems where qubits are addressed in frequency. 
\end{abstract}

\maketitle

\section{Introduction}

Fast and high-fidelity manipulation of qubit states is one of the primary requirements for fault tolerant quantum information processing and quantum computing. Actually realizing high-fidelity on arbitrary states are in the best cases limited by fundamental quantities such as coherence time or available bandwidth due to level separations in atomic systems. At the same time, in a fully fault tolerant quantum algorithm, the large majority of the qubits are used by the error correction \cite{Fowler2012}, where these ancilla qubits have to be read out and re-initialized on the fly repeatedly during the sequence. For initialization operations, the input state is not arbitrary but well known, based on the knowledge of the initial state, shortcut to adiabatic pulses can be used to create faster and more robust initialization pulses. High-fidelity initialization on qubits that are closely spaced in frequency generally requires: (i) a short operation time compared to the coherence time, (ii) robustness against imperfections in the system, for example, frequency variations, frequency detunings, and Rabi frequency fluctuations, (iii) reasonably low off-resonant excitations of frequency-neighboring qubits, which is a practical obstacle for high-fidelity manipulation on such qubits. For example, the frequency variation when making superconducting transmon qubits is typically $>$2\%. Further, when coupling several transmon qubits, two-qubit operation frequencies will be relatively close in frequency compared to the typical Rabi frequencies of 20-100 MHz, putting large demands on the frequency selectivity of the gate operations \cite{Jonas2018,Nicolas2018}. Another example is the ensemble rare-earth ions (REI) system, where the qubit is represented by an ensemble of ions with hundreds-kHz inhomogeneous broadening and different qubits are closely spaced in frequency \cite{Ohlsson2002,Lars2008}. The requirements on low off-resonant excitation originates from the fact that qubits at neighboring frequency channels could be excited off-resonantly by the pulses targeting the qubit ions, and disturb the qubit operation resulting in a reduction of operational fidelity.

Combining speed and robustness is challenging, as some pulses are good at one aspect but not the other. For example, hard resonant pulses are fast but highly sensitive to frequency detunings and fluctuations in light intensity; adiabatic-passage pulses (both rapid and stimulated adiabatic passage) are robust against variations but relatively slow \cite{Bergmann1998, Vitanov2001, Kral2007}; composite pulses are robust benefitting from the cancellation of errors by a small number of carefully-designed, consecutively-implemented rotations, but involve multiple pulses so the operation time might be long \cite{Levitt1986, BT2011}. Protocols that can be used to generate pulses which are good at both aspects are optimal control theory and shortcut to adiabaticity (STA). Optimal control theory can be used to optimize one or more physical quantities to achieve pulses that enable fast operation and are immune to experimental errors \cite{Malinovsky1997, Vasilev2009, Andreas2009, KKobzar2012}. Smooth optimal control has been employed in nitrogen-vacancy centers in diamond to achieve robustness with respect to variation in the control field and frequency detuning from inhomogeneous broadening \cite{Tobias2015}. A robust NOT gate has been proposed where a hybrid protocol involving both inverse engineering in STA and optimal control was used \cite{Leo2017}. STA intends to speed up the adiabatic process via nonadiabatic dynamics while retaining the robustness achieved by the adiabatic process \cite{Xchen2010b, Xchen2012}, for this several techniques have been developed, including inverse engineering based on Lewis-Riesenfeld (LR) invariants \cite{Xchen2012,Xchen2011}, counter-diabatic driving (or transitionless driving) \cite{Berry2009, Xchen2010}, and fast-forward technique \cite{Masuda2010}. In the inverse engineering technique based on LR invariants, the desired target state is guaranteed by the boundary constraints on the instantaneous qubit state, which is the eigenstate of the invariant and analytically known. The freedom left provides the flexibility to optimize 
the parameters of pulse with respect to the imperfections in the system under consideration \cite{NJP2012,PRLStephane}. Nevertheless, counter-diabatic driving can steer the qubit state nonadiabatically along the adiabatic path by adding an additional term to the system Hamiltonian. The physical implementation of this term could be a microwave between two qubit levels or modifications on the light pulses through unitary transformations \cite{DuYX2016,YiChaoPRA,Alexandre2016,Brain2017, Henrik2018}. These two techniques are mathematically equivalent \cite{Xchen2011} but are different in terms of physical implementation \cite{Ban2018}. So for a specific system one might be more applicable than the other depending on the physical constraints and limitations.

In this work, we theorectically propose a protocol for designing pulses that can manipulate qubits closely spaced in frequency in a three-level system between two arbitrary (but known) states with a high fidelity. A combination of the inverse engineering technique and optimization of pulse parameters is used to develop nonadiabatic two-color resonant STA pulse, which are fast in time, and robust against detuning in frequencies and fluctuations in Rabi frequencies. The protocol is applied in simulation to REI system, which is a competitive approach for quantum computing and quantum memories due to their excellent optical and spin coherent properties. The rare-earth quantum computing scheme is also supported in the fundamental science part of the European quantum technologies flagship program that was initiated in 2018 \cite{QFlagship}. The qubit coherence time can be as long as 6 hours \cite{Zhong2015}, and the optical coherence time can be several milliseconds \cite{Equall1994}. In this approach REI are doped into a crystal, which acts as a natural trap, trapping the ions with sub-nm separation. The high qubit density offers ideal potential for scaling, and several promising schemes towards scalable quantum computers have been suggested \cite{Ohlsson2002,Andreas2009,Andreas2015}. Using the inverse engineering technique, robust nonadiabatic pulses are developed for the REI system, which can reduce the qubit initialization time from the previous value of 17.6 $\mu$s using the adiabatic complex hyperbolic secant (CHS) pulses to 4 $\mu$s with realistic Rabi frequencies \cite{Lars2008,Roos2004}. Most importantly, the time that the ions spend in the excited state with our pulses is only 0.7 $\mu$s, which is reduced by a factor of 6 compared to the previous value in the same system. This would give a corresponding factor of 6 reduction in the fidelity error in coherence time ($T_2$) limited systems. The combination of inverse engineering with optimization of pulse parameters may be used to tailor the performance of light-matter interaction in other frequency addressing systems such as superconducting qubits, and may also be used for optimizing the bandpass feature around a center frequency in waveguides \cite{Koushik2015, Tseng2012}.

The article is organized as follows. In Section 2 the Hamiltonian of a three-level system and its LR invariant are described. Design of the pulses utilizing the inverse engineering technique, as well as the performance of the pulses are presented in Section 3, where application examples of the protocol for other operation tasks are also provided. Discussion and conclusion on the protocol is made in Section 4. Optimization method of the pulse parameters and a summary of all pulses presented in this work can be found in Appendix A and B, respectively.

\section{Hamiltonian and Lewis-Riesenfeld invariants}

In a laser adapted interaction picture and within the rotating wave approximation, we write the Hamiltonian of a three-level system, as shown in Fig. 1, in stimulated Raman adiabatic passage for a ``one-photon resonance'' case in bases of $\left|1\right>$, $\left|\rm{e}\right>$, and $\left|0\right>$ as \cite{Xchen2012, Bergmann1997}
\begin{equation}
H(t) = \frac{\hbar}{2}\begin{bmatrix}  
0 & \Omega_{\rm{p}}(t) & 0 \\
\Omega_{\rm{p}}(t) & 0 & \Omega_{\rm{s}}(t)e^{-i\varphi}\\
0 & \Omega_{\rm{s}}(t)e^{i\varphi} & 0
\end{bmatrix}.
\end{equation}
$\Omega_{\rm{i}} ~ (\rm{i} = \rm{p,s})$ is the Rabi frequency and represents the coupling between the laser light and the optical transitions. $\varphi$ is a time-independent phase of the field $\Omega_{\rm{s}}$, and serves the purpose of achieving a phase relationship between $\left|0\right>$ and $\left|1\right>$ depending on the target state.

\begin{figure}[htbp] 
\centering
\includegraphics[width=3.5cm,height=5cm]{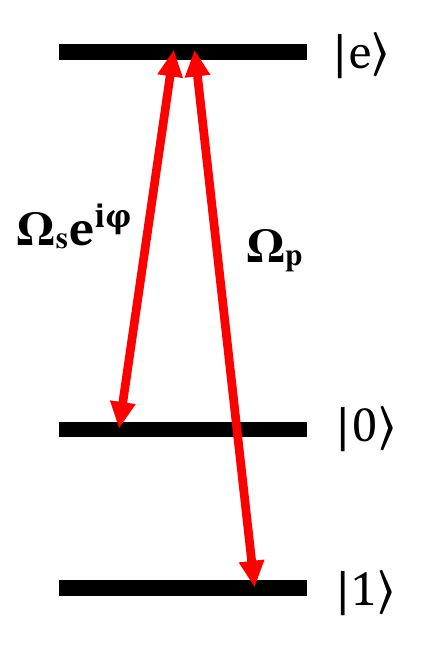}
\caption{Schematic energy levels of a three-level $\Lambda$ system. The qubit is represented by two long-lived ground state levels $\left|0\right>$ and $\left|1\right>$, where $\left|1\right>$ is initially populated. Qubit levels are coupled through optical transitions $\left|0\right>-\left|\rm{e}\right>$ and $\left|1\right>-\left|\rm{e}\right>$, which possibly exhibits an inhomogeneous broadening. $\Omega_{\rm{p}}$ and $\Omega_{\rm{s}}$ denote the respective Rabi frequencies. $\varphi$ is a time-independent phase factor of $\Omega_{\rm{s}}$. }
\label{fig.1}
\end{figure}

In the following we employ the LR invariants theory to find the solution of the Schr\"odinger equation \cite{Lewis1969}

\begin{equation}
i\hbar \partial_t \left|\psi(t)\right> = H(t) \left|\psi(t)\right>.
\end{equation}
The LR invariant $I(t)$ of $H(t)$, which satisfies $dI/dt \equiv \partial I/\partial t + (1/i\hbar)[I(t), H(t)] = 0 $, can be constructed as \cite{Xchen2011, YZLai1996}
\begin{equation}
I(t)=\frac{\hbar\Omega_0}{2}\begin{bmatrix}
0 & \cos\gamma \sin\beta & -i\sin\gamma e^{-i\varphi} \\
\cos\gamma\sin\beta & 0 & \cos\gamma\cos\beta e^{-i\varphi}\\
i\sin\gamma e^{i\varphi} & \cos\gamma\cos\beta e^{i\varphi} & 0
\end{bmatrix},
\end{equation} where $\Omega_0$ is an arbitrary constant in unit of frequency. $\gamma$ and $\beta$ are time-dependent variables to be designed, and relate to the Rabi frequencies as:
\begin{equation}
\Omega_{\rm{p}}(t) = 2[\dot{\beta}\cot{\gamma(t)}\sin{\beta(t)}+\dot{\gamma}\cos{\beta(t)}],
\end{equation}
\begin{equation}
\Omega_{\rm{s}}(t) = 2[\dot{\beta}\cot{\gamma(t)}\cos{\beta(t)}-\dot{\gamma}\sin{\beta(t)}].
\end{equation} In these equations $\dot{\beta}$ and $\dot{\gamma}$ denote the time derivative of $\beta$ and $\gamma$, respectively.
The eigenstates of the invariant $I(t)$ are
\begin{equation}
\left|\phi_0(t)\right>=\begin{bmatrix}
    \cos\gamma(t)\cos\beta(t)\\
    -i\sin\gamma(t)\\
    -\cos\gamma(t)\sin\beta(t)e^{i\varphi}
    \end{bmatrix}
\end{equation}
and
\begin{equation}
\left|\phi_\pm(t)\right>= \\\\
 \frac{1}{\sqrt{2}}\begin{bmatrix}
    \sin\gamma(t)\cos\beta(t) \pm i\sin\beta(t)\\
    i\cos\gamma(t)\\
    [-\sin\gamma(t)\sin\beta(t) \pm i\cos\beta(t)]e^{i\varphi}
    \end{bmatrix},
\end{equation} with eigenvalues $\lambda_0 = 0$, and $\lambda_\pm = \pm\hbar\Omega_0/2$, respectively.

The LR theory tells that any linear summation
\begin{equation}
\left|\psi(t)\right>= \sum_{n = 0, \pm} C_n e^{i\alpha_n} \left|\phi_n(t)\right>,
\end{equation}
where $C_n$ is a time-independent amplitude determined by boundary conditions and 
$$\alpha_n(t) = \frac{1}{\hbar}\int_0^t \left< \phi_n(t')|i\hbar \frac{\partial}{\partial t'}-H(t')| \phi_n(t')\right> dt',$$ 
is the LR phase, is a solution of Eq. (2). From Eqs. (1) and (6) one can prove that $\alpha_0 = 0$, which means that $\left|\phi_0(t)\right>$ is a solution to Eq. (2) as well.

In this work, we focus on the case where $H(t)$ in Eq. (1) drives the 3-level system from an initial state, for example $\left|1\right>$, to an arbitrary superposition target state $\left|\psi_{\rm{tg}}\right> = \cos\theta_a \left|1\right> + \sin\theta_a e^{i\varphi_a} \left|0\right>$ ($\theta_a$ and $\varphi_a$ are arbitrary angles) or vice versa, along the invariant eigenstate $\left|\phi_0(t)\right>$.

\section{Inverse engineering and results}

Boundary states $\left|\phi_0(0)\right> = \left|1\right>$ and $\left|\phi_0(t_{\rm{f}})\right> = \left|\psi_{\rm{tg}}\right>$ ($t_{\rm{f}}$ refers to the ending time of the laser pulse) impose restrictions on the boundary values of $\gamma(t)$ and $\beta(t)$. In the simplest form but without loss of generality, one may choose

\begin{equation}
\left\{
\begin{array}{rcl}
&\gamma(0) = {0},& \gamma(t_{\rm{f}}) = \pi \\
&\beta(0) ={0},& \beta(t_{\rm{f}}) = \pi-\theta_a,\\
\end{array} \right.
\end{equation}
and $\varphi = \varphi_a$.

Bearing these boundary values in Eq. (9) in mind, and also considering the requirements on both the robustness and off-resonant excitations as described in the Introduction, we make an ansatz on $\gamma(t)$, which fulfills Eq. (9) and reads
\begin{equation}
\gamma(t) = \frac{\pi}{t_{\rm{f}}} t + \sum_{n=1}^{\infty} a_n \sin\left(\frac{n\pi}{t_{\rm{f}}} t\right),
\end{equation}
where $a_n$ is the coefficient of each sinusoidal component, which will be discussed shortly.

We further let $\dot{\beta}(t)\propto \sin(\gamma)$ to fix the divergence of $\Omega_{\rm{p,s}}$ at $t = 0$ and $t = t_{\rm{f}}$ in Eqs. (4) and (5). The ansatz on $\beta(t)$, which satisfies Eq. (9), reads
\begin{equation}
\beta(t) = \frac{\pi-\theta_a}{2} [1-\cos\gamma(t)].
\end{equation}
With Eqs. (10) and (11), Rabi frequencies in Eqs. (4) and (5) are simplified as:
\begin{equation}
\Omega_{\rm{p}} = \dot{\gamma}(t)[(\pi-\theta_a)\cos{\gamma(t)}\sin{\beta(t)}+2\cos{\beta(t)}],
\end{equation}
\begin{equation}
\Omega_{\rm{s}} = \dot{\gamma}(t)[(\pi-\theta_a)\cos{\gamma(t)}\cos{\beta(t)}-2\sin{\beta(t)}].
\end{equation}

The ansatzs shown in Eqs. (10) and (11) ensure that $H(t)$ drives the system from $\left|1\right>$ to $\left|\psi_{\rm{tg}}\right>$. However, from an experimental point of view $\Omega_{\rm{p,s}}$ is preferred to start and end at zero, that is
\begin{equation}
\Omega_{\rm{p,s}}(0)=\Omega_{\rm{p,s}}(t_{\rm{f}})=0,  
\end{equation}
equivalently
\begin{equation}
\dot{\gamma}(0)=\dot{\gamma}(t_{\rm{f}})=0,  
\end{equation}
so as to avoid sharp changes in electric field at $t = 0$ and $t = t_{\rm{f}}$. This is because sharp changes in time domain would result in multiple frequency components in frequency domain, which could cause unwanted excitations. Eq. (15) requires
\begin{equation}
a_1+3a_3+5a_5+7a_7 + 9a_9 + ... + (2k-1)\cdot a_{2k-1} = 0
\end{equation} and
\begin{equation}
a_2+2a_4+3a_6+4a_8+ ... + k\cdot a_{2k} = -0.5,
\end{equation}
where $k$ is an integer number. In principle, $k$ can be infinitely large, but in practice it will be limited by the rise time of the device which is used for generating the pulses, e.g. an acousto-optical modulator. In this work, $k=4$ is considered.

Irrespective of the values of $a_n$ as long as they satisfy Eqs. (16) and (17), pulses constructed from $\gamma(t)$ and $\beta(t)$ from Eqs. (10) and (11) are able to drive the qubit from $\left|1\right>$ to $\left|\psi_{\rm{tg}}\right>$. However, the fidelity might not be robust. The 6 degrees of freedom available in Eqs. (16) and (17) can be used to tailor the pulses for achieving the required robustness. For simplicity, in this work we consider the case where $a_{1,3,5,7} = 0$, and $a_{2,6,8}$ are to be optimized. The optimization is done by manually optimizing the parameters one at a time. Details can be found in Appendix A. While even better results could be expected by using all 6 degrees of freedom, the present more restricted choice still clearly demonstrates the strength of the approach.

Below we apply the pulse-designing protocol discussed above to an ensemble REI qubit system to develope fast and robust initialization pulses. In this system, the qubit is represented by two hyperfine ground states of an ensemble of REI randomly doped into an \YSO crystal, for example, Pr$^{3+}$:\YSO or Eu$^{3+}$:\YSO. Experimental arbitrary gates have been demonstrated in the Pr system \cite{Lars2008}, making it a useful choice for comparing our results with. The relevant energy structure is shown in Fig. 2(a). The qubit operations between $\left|0\right>$ and $\left|1\right>$ are implemented via the respective optical transitions $\left|0\right>-\left|\rm{e}\right>$ and $\left|1\right>-\left|\rm{e}\right>$ at a wavelength of 605.977 nm with an optical coherence time of about 152 $\mu$s (in a magnetic field of 77 G) \cite{Equall1995}. The qubit ions can be spectrally isolated in a zero-absorption frequency range of 18 MHz (called a spectral pit hereafter) and have an inhomogenous full width at half maximum (FWHM) frequency span of 170 kHz \cite{Lars2008}. A schematic of the absorption spectrum of a qubit in a pit is shown in Fig. 2(b), where both $\left|0\right>$ and $\left|1\right>$ are populated, and zero frequency is defined as the transition frequency from $\left|0\right>$ to the first level of the excited states. Population of $\left|0\right>$ ($\left|1\right>$) state is represented by peaks 1-3 (peaks 4-5) in the absorption spectrum separated by the upper state splitting, 4.6 MHz and 4.8 MHz, where the different peak heights originate from different transition oscillator strengths. The distance from peak 2 or peak 5 to the edge of the spectral pit is about 3.9 MHz. Here 3.5 MHz can be considered leaving a margin of 0.4 MHz accounting for the width of the peak. High fidelity qubit manipulation in such a system requires that the action on the system by the pulses to be flat within a frequency detuning range of at least $\pm$ 170 kHz, and leave the ions which are 3.5 MHz away from the qubit ions in frequency domain untouched.

\begin{figure} [htbp]
	\begin{minipage}{8cm}
\centering
\includegraphics[width=4.6cm,height=6cm]{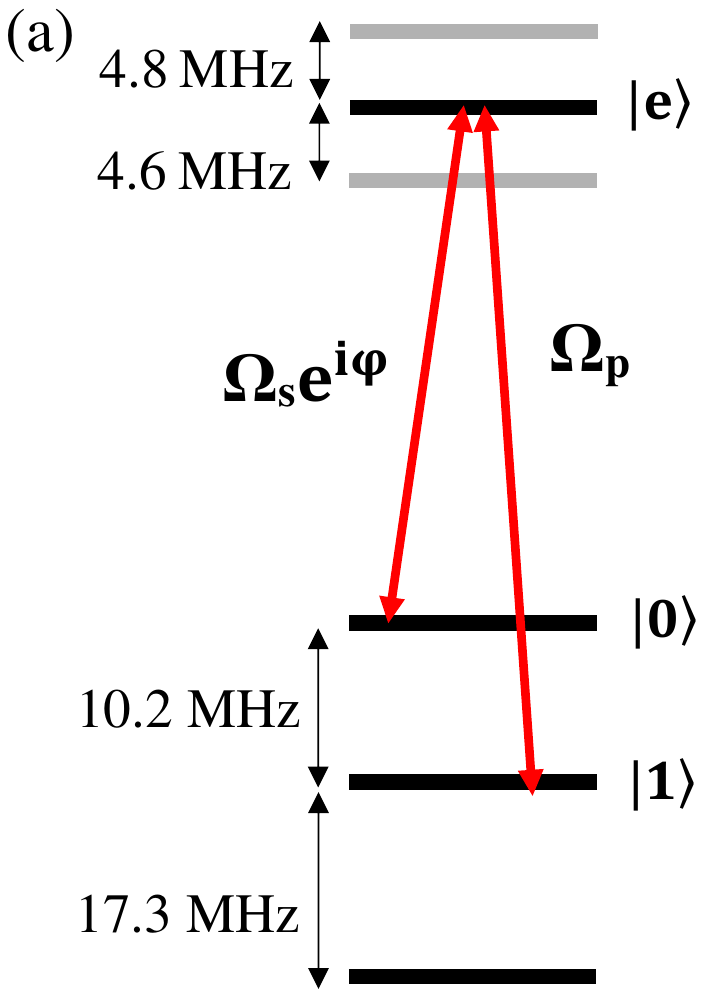}
\end{minipage}
\hfill  
\begin{minipage}{8cm}
\includegraphics[width=8cm,height=4.2cm]{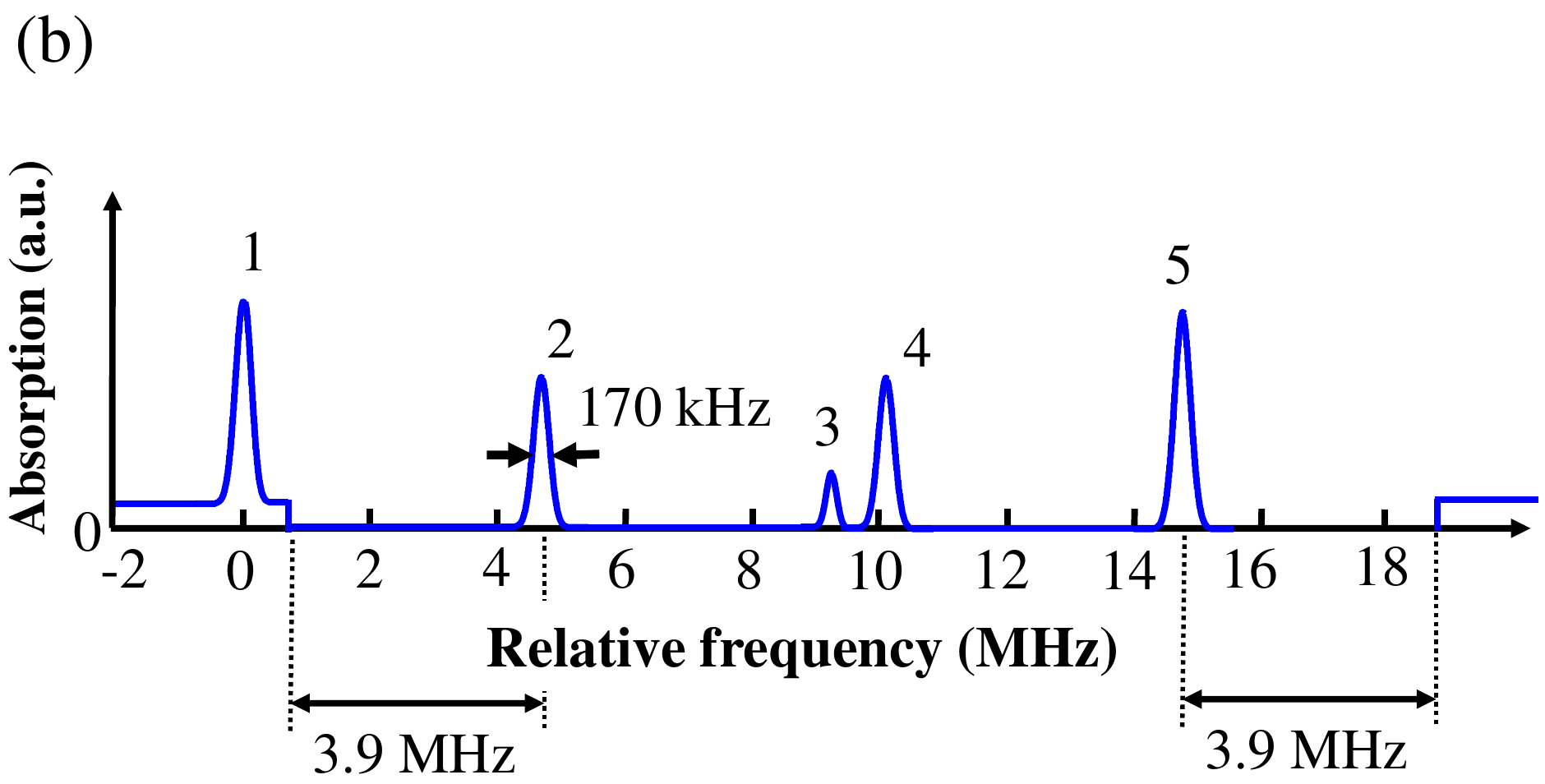}
\end{minipage}
\label{fig.2}
\caption{(a) The relevant energy levels of Pr ions in an \YSO crystal. The qubit is represented by two ground state levels $\left|0\right>$ and $\left|1\right>$, where $\left|1\right>$ is initially populated. Qubit levels are coupled through optical transitions $\left|0\right>-\left|\rm{e}\right>$ and $\left|1\right>-\left|\rm{e}\right>$, both of which have an inhomogeneous FWHM linewidth of 170 kHz. (b) A schematic diagram of the absorption spectrum of a qubit in a zero absorption spectral window. Peak 1-3 represent absorption from $\left|0\right>$ to each level of the excited states, and peak 4-5 from $\left|1\right>$ to the two lower levels in the excited states, respectively. The distance from peak 2 or peak 5 to the edges of the pit is 3.9 MHz.}
\end{figure}

In the following passage we set $\left|\psi_{\rm{tg}}\right> = \frac{1}{\sqrt{2}} (\left|1\right> + \rm{i}\left|0\right>)$ as an example, to demonstrate the pulse performance in the ensemble REI system. The pulse duration is set to $t_{\rm{f}} = 4$ $\mu$s for all cases shown in this work, which is restricted by practical limitations such as the maximally available intantaneous Rabi frequencies, and maximal rising speed in Rabi frequency. The optimized $a_n$ parameters are: $a_2 = -1.10, a_6 = 0.06$, and $a_8 = 0.02$. Dynamics of $\Omega_{\rm{p,s}}$ in Eqs. (12) and (13) as well as the state evolutions are shown in Fig. 3(a) and 3(b), respectively. The maximal instantaneous Rabi frequencies are less than 1.6 MHz for both fields, which is a realistic value for current ensemble experiment implementations. The final population shown in Fig. 3(b) is equally distributed in $\left|1\right>$ and $\left|0\right>$, while $\left|\rm{e}\right>$ is unpopulated as expected. Within the operation period, the average time that ions spent in the excited state is 0.7 $\mu$s, which is reduced by a factor of 6 compared to the previous value obtained with CHS pulses \cite{Lars2008}. This reduction would clearly improve the operational fidelity for Pr$^{3+}$ qubits, where $T_2$ is 50 $\mu$s for the excitation density of $3\times10^{14}/\rm{cm^3}$ used in reference \cite{Lars2008}. The relatively high ratio between the time spent in the excited state and $T_2$ was concluded to be the primary limitation to the fidelity in experiment \cite{Lars2008}.

In the following, the performance of the pulses shown in Fig. 3(a) is evaluated from two aspects: robustness of fidelity against frequency detuning around the center (section 3.1); and the off-resonant excitations on ions which sit $>$3.5 MHz away from the qubit (section 3.2).

\subsection{Ultra-robust fidelity}

Fidelity of achieving the arbitrary single qubit state (ASQS) $\left|\psi_{\rm{tg}}\right>$ is calculated as
\begin{equation}
F = \left|\left<\psi_{\rm{tg}}|\psi(t_{\rm{f}})\right>\right|^2,
\end{equation}
where $\left|\psi(t_{\rm{f}})\right> = [C_{1}(t_{\rm{f}}), C_{\rm{e}}(t_{\rm{f}}), C_{0}(t_{\rm{f}})]^T$ denotes the state of qubit at $t = t_{\rm{f}}$. $C_\textrm{n} (t_{\rm{f}})$ (n=1, $\rm{e}$, 0) is the probability amplitude of the final state, which is the numerical solution of the three-level coupled differential equations. These equations originate from the Schr\"odinger equation and read
\begin{equation}
\begin{bmatrix}
\dot{C}_1\\
\dot{C}_{\rm{e}}\\
\dot{C}_0
\end{bmatrix}
= -\frac{i}{2}\begin{bmatrix}
0 & \Omega_{\rm{p}} & 0\\
\Omega_{\rm{p}} & 2\Delta & \Omega_{\rm{s}} e^{-i\varphi_a}\\
0 & \Omega_{\rm{s}} e^{i\varphi_a} & 0
\end{bmatrix}
\cdot \begin{bmatrix}
C_1\\
C_{\rm{e}}\\
C_0
\end{bmatrix}
\end{equation}
where $\Delta$ represents the detuning between the individual transition frequencies of the inhomogeneously broadened qubit ions and the center frequency of the pulses.

\begin{figure}[htbp]
\centering
\begin{minipage}{8cm}
\centering
	\includegraphics[width=6.5cm,height=4.5cm]{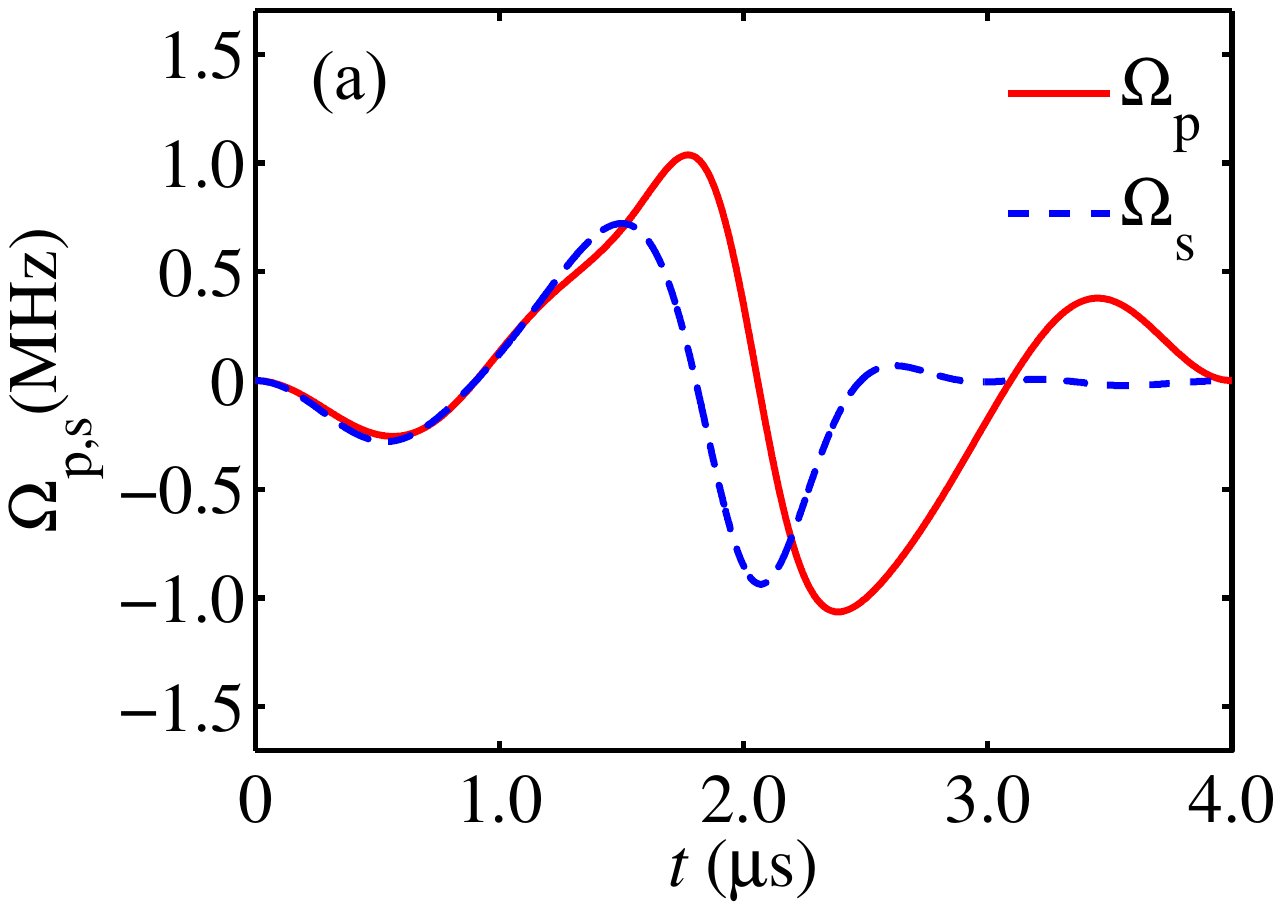}
\end{minipage}
\hfill  
\begin{minipage}{8cm}
\centering
\includegraphics[width=6.2cm,height=4.5cm]{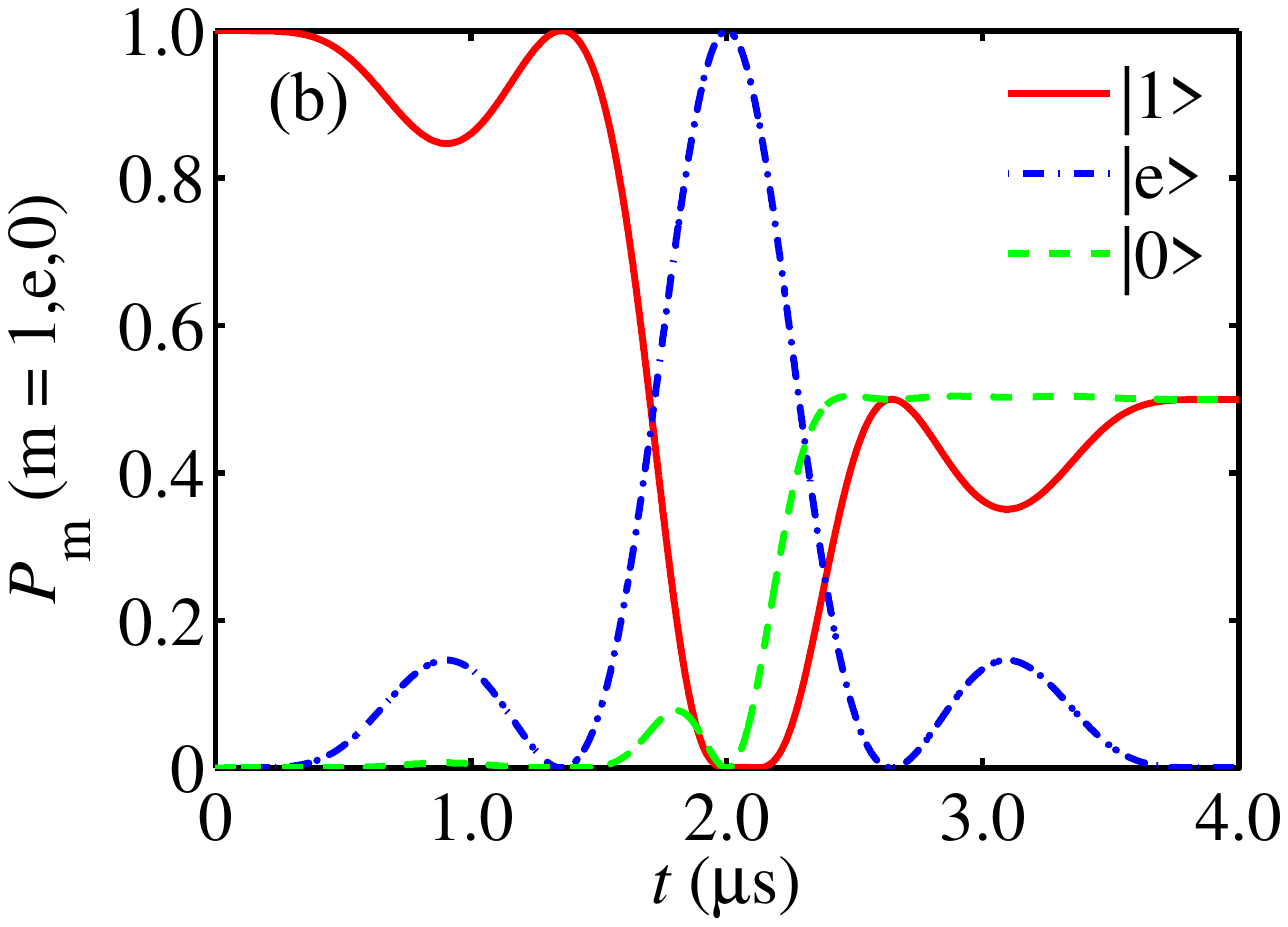}
\end{minipage}
	\caption{(a) Time dependence of Rabi frequencies, where the solid-red (dashed-blue) line denotes $\Omega_{\rm{p}}$ ($\Omega_{\rm{s}}$). (b) Time evolution of the population on level $\left|1\right>$ ( Solid-red line), $\left|\rm{e}\right>$ (dash-dotted-blue line), and $\left|0\right>$ (dashed-green line). $a_2 = -1.10$, $a_6 = 0.06$ and $a_8 = 0.02$, which are optimized to achieve high robustness against frequency detuning and minimize the off-resonant excitations, and $\left|\psi_{\rm{tg}}\right> = \frac{1}{\sqrt{2}} (\left|1\right> + \rm{i}\left|0\right>)$.}
\label{fig.3}
\end{figure} 

The dependence of $F$ in Eq. (18) on frequency detuning $\Delta$, using the optimized $a_n$ parameters mentioned above, is shown as the solid-red curve in Fig. 4. Within $\pm$340 kHz around the center frequency the average fidelity is 99.8$\%$. Above $\pm$3.5 MHz the fidelity stabilizes to 50 $\%$, which is because the ions at that spectral range still remain in the initial state of $\left|1\right>$ as they are hardly excited (shown later in Fig. 6). The feature inbetween $\pm$[1, 3.5] MHz frequency detuning does not matter, as there are no ions there because they have been removed by optical pumping. In comparison, the fidelity achieved by the CHS pulses in simulation is shown as the blue-dashed curve (the parameters used here are the same as those in \cite{Lars2008}), where the average fidelity over $\pm$340 kHz ($\pm$170 kHz) is 86.9\% (98.3\%). Clearly the robustness in fidelity of shortcut pulse is improved, and the pulse duration is only one quarter of that of the CHS pulses.

\begin{figure}[htbp]
\centering
\includegraphics[width= 7 cm,height=5cm]{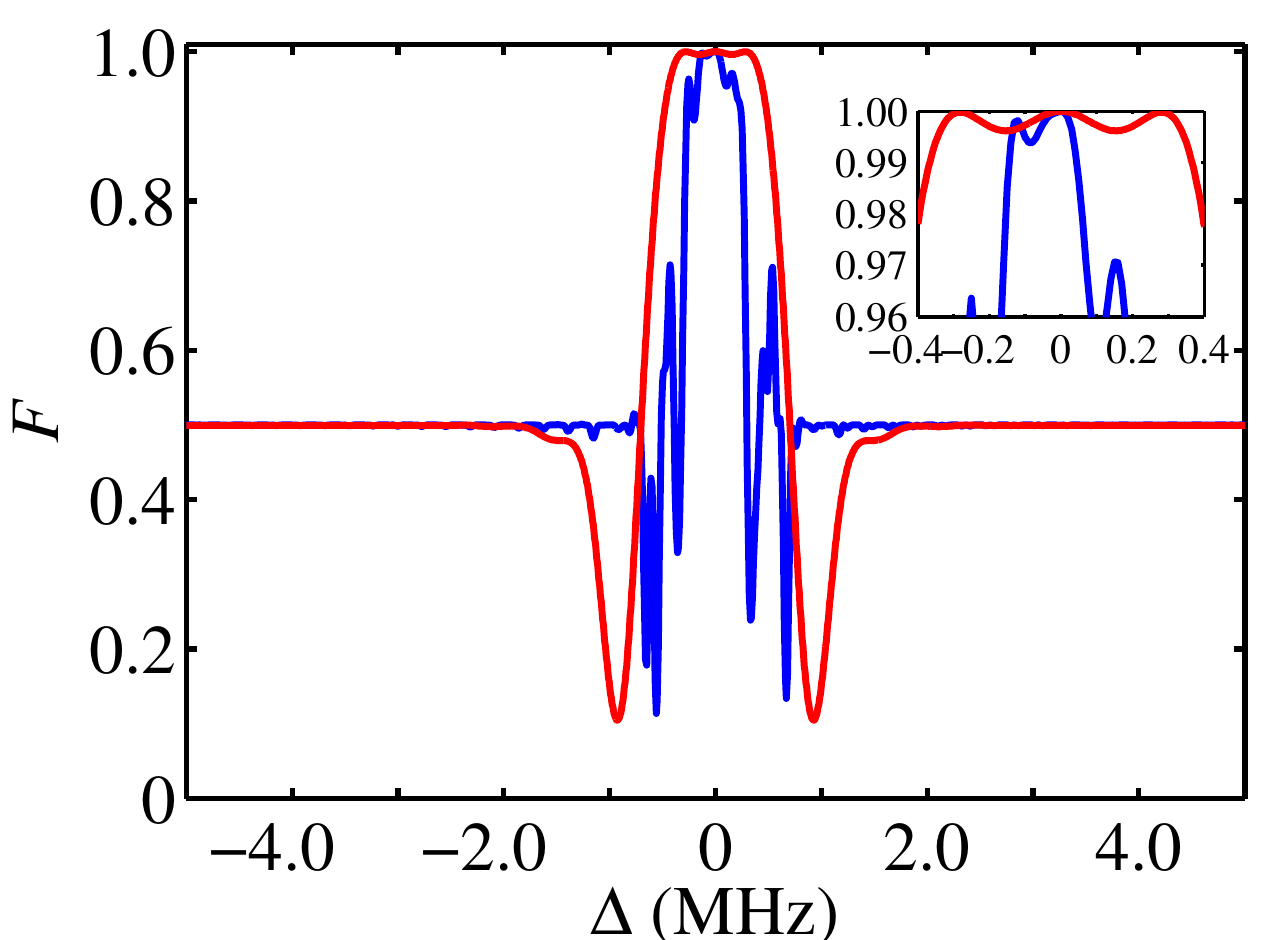}
\caption{Dependence of fidelity of achieving $\left|\psi_{\rm{tg}}\right> = \frac{1}{\sqrt{2}} (\left|1\right> + \rm{i}\left|0\right>)$ on frequency detuning. Solid-red line: fidelity achieved with the shortcut pulses developed in this work with optimized parameters ($a_2 = -1.10$, $a_6 = 0.06$ and $a_8 = 0.02$). Solid-blue line: the fidelity achieved by the complex hyperbolic secant pulses. The insert is a magnification of the center frequency range.}
\label{fig.4}
\end{figure} 

Robustness of fidelity in response to fluctuations in both $\Omega_{\rm{p}}$ and $\Omega_{\rm{s}}$ (assume both fields are generated from the same light source) for the shortcut pulses is shown in Fig. 5(a), where blue-dashed (solid-red) line shows the dependence under condition of no detuning (with 170 kHz detuning), and $\eta$ describes the relative change in Rabi frequencies, i.e. $\eta= \Delta\Omega_{\rm{p,s}}/\Omega_{\rm{p,s}}$. The dependence does not change appreciatively in response to different detunings. In both cases, fidelity is much less sensitive on positive fluctuations than the negative ones, This indicates that it is safer to have higher Rabi frequencies than lower ones in the implementation of the pulses.  The dependence of using CHS pulses is shown in Fig. 5(b), where fidelity is very robust against Rabi fluctuations if there is no detuning (dashed-green). However, fidelity drops clearly in the presence of detuning (solid-purple) and there is no significant difference in sensitivity between positive and negative fluctuations.

\begin{figure}[htbp]
\begin{minipage}{8cm}
\centering
\includegraphics[width=6.5cm,height=4.5cm]{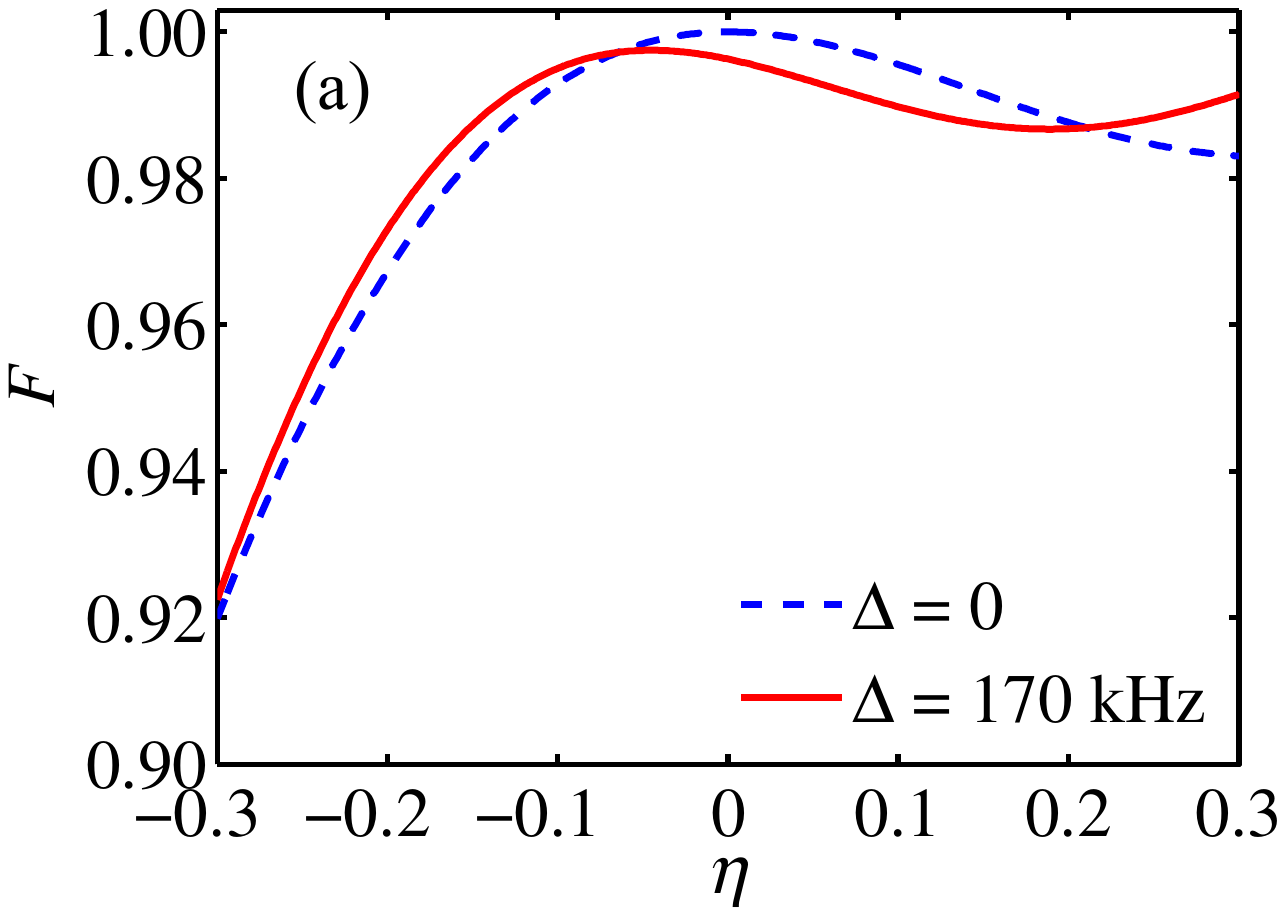}
\end{minipage}
\hfill  
\begin{minipage}{8cm}
\centering
\includegraphics[width=6.5cm,height=4.5cm]{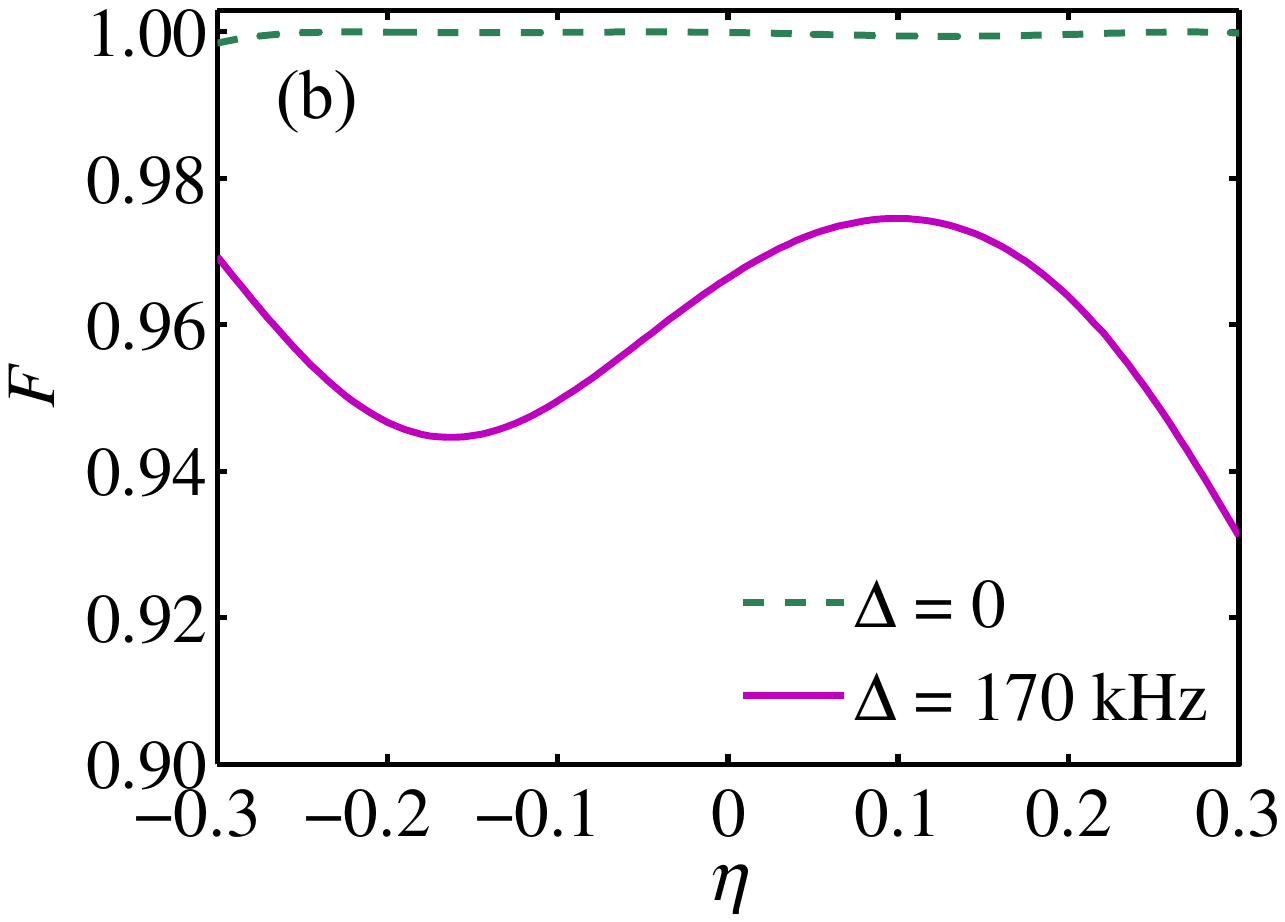}
\end{minipage}
\caption{Dependence of fidelity of achieving $\left|\psi_{\rm{tg}}\right> = \frac{1}{\sqrt{2}} (\left|1\right> + \rm{i}\left|0\right>)$ on the fluctuations in Rabi frequencies for shortcut pulses (a) and complex hyperbolic secant pulses (b), where $\eta$ describes the relative change in $\Omega_{\rm{p,s}}$. Blue-dashed and green-dashed curves show the dependences under conditions of no detuning, and solid-red and solid-purple for the cases where detuning is 170 kHz.}
\label{fig5}
\end{figure}

\subsection{Low off-resonant excitations}

Besides the robustness against frequency detuning, high-fidelity qubit operations in ensemble REI system also require low off-resonant excitations on the ions that are spectrally close to the qubit ions. In the ideal case, these ions should not be affected by the light pulses targeting the qubit ions, but rather remain in their initial states. Fig. 6(a) shows the population of the final state $\left|\psi(t_{\rm{f}})\right>$ in levels $\left|1\right>$, $\left|\rm{e}\right>$ and $\left|0\right>$ as function of detuning frequencies with the same optimized $a_n$ values as in Section 3.1. At far off-resonant frequencies ($>$5 MHz), the ions indeed remain in the initial state $\left|1\right>$. For those ions with detuning above 3.5 MHz, less than 2$\%$ are transferred to state $\left|0\right>$ by the light pulses. While this is not perfect, it might be acceptable as the ion density in this interval is less than that of the qubit ions, and can be further reduced as the robustness region over frequency detuning does not need to be as large as $\pm$340 kHz, considering the $\pm$170 kHz Gaussian frequency distribution of the ensemble ions in the qubit. As a comparison, using the CHS pulses the off-resonant excitation is below 2\% for detuning above 1.2 MHz, and nearly zero at 3.5 MHz, which is better than the shortcut pulses. The conceivable reason for this may be that the spectral coverage of the shortcut pulses is larger as Rabi frequencies change with time more rapidly.

\begin{figure}[htbp]
\centering
\includegraphics[width=7cm,height=5cm]{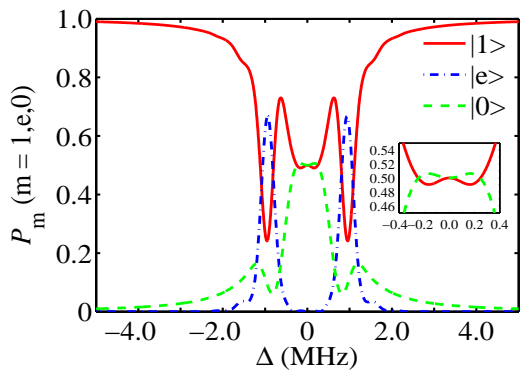}
\label{fig.6}
\caption{Population of final state $\left|\psi(t_{\rm{f}})\right>$ in level $\left|1\right>$ (solid-red), $\left|\rm{e}\right>$ (dash-dotted-blue), and $\left|0\right>$ (dashed-green) as a function of detuning frequencies with optimized $a_n$ values. The off-resonant excitation at 3.5 MHz is less than 2.0\%, which can be further reduced by sacrificing the width of the robust region in the center. }
\end{figure}

\subsection{Application examples for other operation tasks}

The protocol discussed above can be applied to any single-qubit operation tasks as long as the initial state is known. Below we provide two examples.

\textbf{(i)} Population transfer from state $\left|1\right>$ to $\left|\rm{e}\right>$ in a two-level system

Let $\beta(t)\equiv0$, the three-level model described in Eqs. (1)-(8) reduces to a two-level model. Rabi frequencies in Eqs. (4) and (5) turns to
\begin{equation}
\Omega_{\rm{p}} = 2\dot{\gamma}(t),
\quad\Omega_{\rm{s}} = 0.
\end{equation}

Eq. (6) tells us that the boundary conditions on $\gamma(t)$ may be
\begin{equation}
\gamma(0) = 0, \quad\gamma(t_{\rm{f}}) = -\pi/2.
\end{equation} We propose an ansatz on $\gamma(t)$ as
\begin{equation}
\gamma(t) = -\frac{\pi}{2t_{\rm{f}}}t + \sum_{n=1}^{2k} a_n \sin\left(\frac{n\pi}{t_{\rm{f}}}t\right),
\end{equation}
where coefficients $a_n$ satisfy 
\begin{equation}
a_2 + 2a_4 + 3a_6 + 4a_8 + ... + k \cdot a_{2k} = 0.25.
\end{equation}
Again considering $t_f$ = 4 $\mu$s and $k = 4$, the optimized parameters are $a_2 = 0.50, a_6 = 0.14,$ and $a_8 = 0$. The pulses can achieve an average fidelity of 99.5$\%$ within $\pm$320 kHz frequency detuning, as shown in Fig. 7(a), which is slightly lower than the fidelity of 99.9\% that CHS pulses can achieve.

Figure 7(b) shows the state evolution (solid-red line) on a Bloch sphere where $\Delta = 170$ kHz. In case of no detuning, the state evolves along the great circle in v-w plane, but at positions around each pole it walks back and forth in a way that is more complicated than with a square pulse.

\begin{figure}[htbp]
\begin{minipage} {8cm}
\centerline{\includegraphics[width=6.5cm,height=4.5cm]{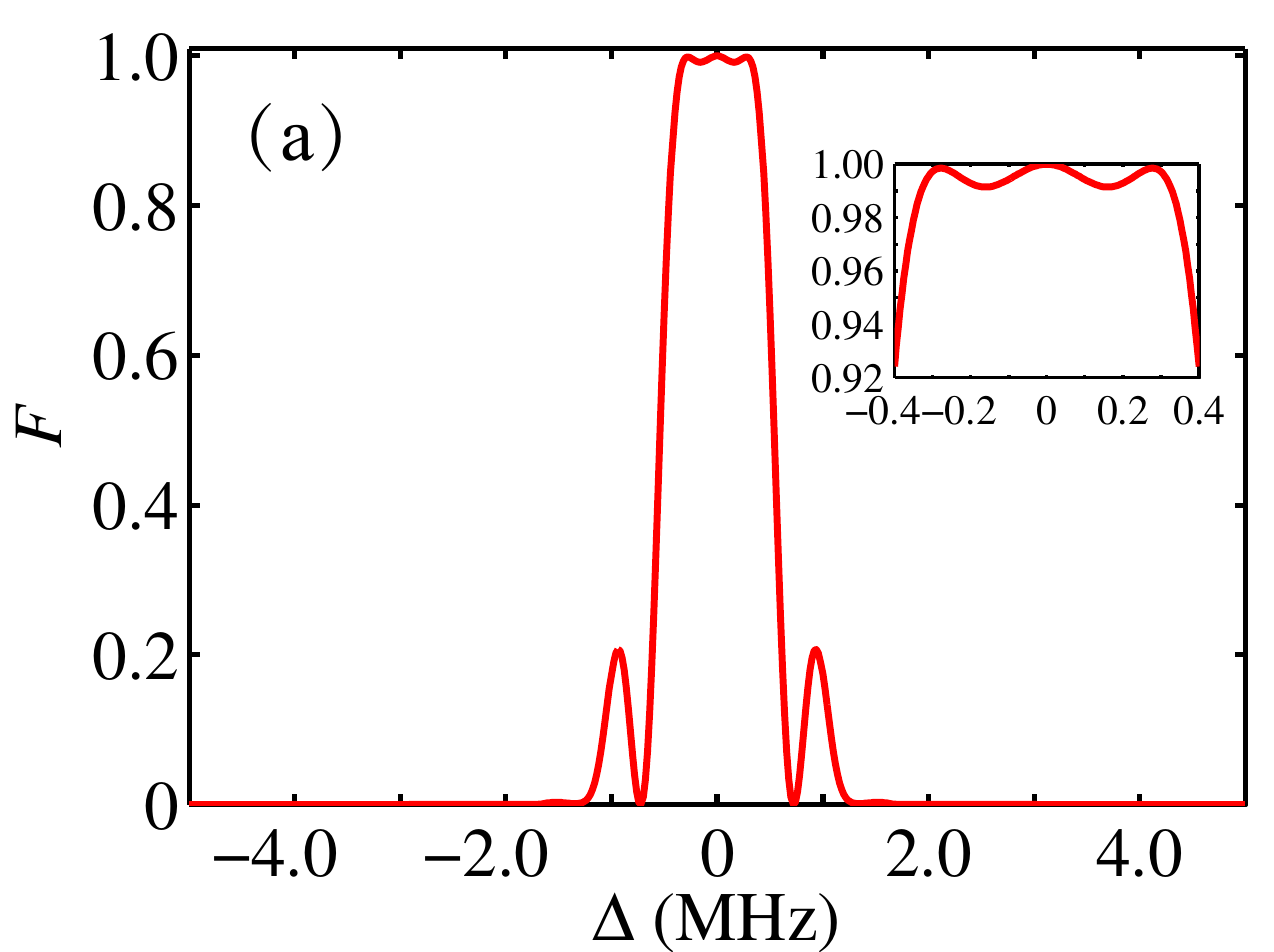}}
\end{minipage}
\hfill
\begin{minipage} {8cm}
\centerline{\includegraphics[width=3.4cm,height=4.0cm]{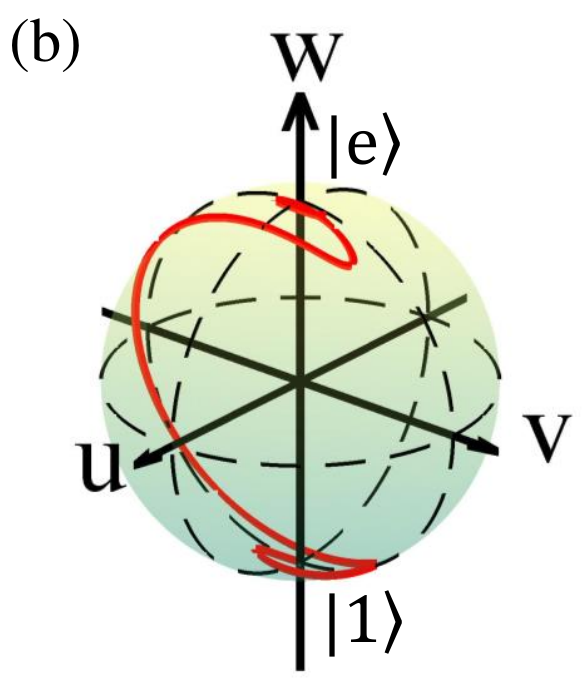}}
\end{minipage}
\caption{(a) Dependence of the fidelity ($F$) for achieving the target state $\rm{i}\left|e\right>$ on frequency detuning $\Delta$. (b) State evolutions (solid-red line) on a Bloch sphere where $\Delta = 170$ kHz. The dashed lines are the three perpendicular great circles in u-v, u-w and v-w plane.}
\label{fig7}
\end{figure}
\textbf{(ii)} Driving an arbitrary superposition state, $\left |\psi_{\rm{i}}\right> = \cos\theta_b \left|1\right> + \sin\theta_b e^{i\varphi_b} \left|0\right>$ ($\theta_b$ and $\varphi_b$ are arbitrary angles in the range of $[0, 2\pi]$), into target state $\left|0\right>$ or $\left|1\right>$.

Here we take the target state $\left|1\right>$ as an illustration. The process is exactly the time reverse of creating an ASQS from state $\left|1\right>$. The boundary conditions in Eq. (9) have to be changed to
\begin{equation}
\left\{
\begin{array}{rcl}
&\gamma(0)& = \pi, \quad\gamma(t_{\rm{f}}) = 0, \\
&\beta(0)& = \pi-\theta_b, \quad\beta(t_{\rm{f}}) = 0,\\
\end{array} \right.
\end{equation}
and $\varphi = \varphi_b$.

Instead of Eq.(10) we assume
\begin{equation}
\gamma(t) = -\frac{\pi}{t_{\rm{f}}} t + \pi + \sum_{n=1}^{2k} a_n \sin\left(\frac{n\pi}{t_{\rm{f}}} t\right),
\end{equation}
and Eqs.(11)-(15) remain unchanged. The required conditions on $a_n$ are as follows:
\begin{equation}
\left\{\begin{array}{rcl}
&a_1+3a_3+5a_5+7a_7+ ...+(2k-1)\cdot a_{2k-1} = 0 &\\
&a_2+2a_4+3a_6+4a_8+...+k\cdot a_{2k} = 0.5.&\\
\end{array} \right.
\end{equation}
where $k$ is an integer. Here we again limit $k =4$ for the same reason as discussed previously.

Taking $\left |\psi_{\rm{i}}\right> = (\left|1\right> + \rm{i} \left|0\right>)/\sqrt{2}$ as an example, the fidelity within $\pm$520 kHz is above 99.9\% with optimized parameters of $a_2 = 1.06, a_6 = 0.16,$ and $a_8 = 0$. Here both the Rabi frequencies and optimized parameters are different from the results in Section 3.1. One can also reuse the pulses presented in Fig. 3(a) by reversing it in time as follows:

\begin{equation}
\Omega^{\rm{new}}_{\rm{p,s}} = -\Omega_{\rm{p,s}}(t_{\rm{f}}-t)
\end{equation}
where $\Omega_{\rm{p,s}}$ are shown as Eq. (12) and (13). The advantage of this option is that one can reuse the $a_n$ values optimized previously (a new optimization might improve the performance slightly).

\section{Discussion and conclusion}

In this article, we propose to combine the inverse engineering based on the LR invariants with the optimization of the multiple degrees of freedom provided by the proper ansatz, to design high-fidelity and experimentally-realistic initialization pulses. The freedom left provides us the flexibility of tailoring the quantum control to be robust against the physical imperfections, in the meanwhile keeping the off-resonant excitations on frequency-neighboring qubits reasonably low. These features, especially the last one, is crucial for achieving high-fidelity manipulation on qubits which are closely spaced in frequency \cite{Roos2004}, for example, superconducting transmon qubits and REI qubits. Instead of frequency detuning and off-resonant excitation, the optimization can also be done for other physical quantities, for example, the time that qubit spend in the excited state, and the lowest maximal Rabi frequencies. As an example, the protocol is used to theoretically derive pulses from a simulation of an ensemble REI system. Nonadiabatic pulses to perform single qubit manipulation between $\left|1\right>$ and an arbitrary superposition state are developed with an operation time of 4 $\mu$s and realistic Rabi frequencies. By optimizing the pulse parameters, simulations show that the average fidelity (assuming an infinite $T_2$) is 99.8\% over a frequency detuning range as large as $\pm$340 kHz, and the off-resonant excitation can be reasonably low. In comparison, in the same system a fidelity of 86.9\% (98.3\%) over $\pm$340 kHz ($\pm$170 kHz) detuning range was achieved by CHS pulses. The significant difference between the shortcut pulses shown in this work and the CHS pulse is that the average time that ions spend in excited state is reduced by a factor of 6. This reduction is important for a decoherence limited system. Considering the experimental dephasing time of 50 $\mu$s ($T_2$) for Pr ions, the fidelity of using our pulses is decreased to 99.1\% whereas it is only 83.2\% using the CHS pulse. Here the fidelity is estimated as $\left<\psi_{\rm{tg}}\right| \rho \left|\psi_{\rm{tg}}\right>$, the density matrix is $\rho = e^{-t_u/T_2} \rho_{\rm{ideal}}+ (1-e^{-t_u/T_2})\rho_{\rm{mixed}}$ \cite{Lars2008}, and $t_u$ = 0.7 $\mu$s. The decoherence only reduces the fidelity by 0.7\% using our pulses, benefiting from a relatively small ratio of $t_u/T_2$. The fidelity are further improved to 99.8\% if qubits ions with long $T_2$ are used, for example Eu$^{3+}$:\YSO, where $T_2$ can be 2.6 ms in a magnetic field of 100 G \cite{Equall1994}.

Comparing with the counter-diabatic driving shortcut protocol\cite{DuYX2016,YiChaoPRA,Alexandre2016,Brain2017}, our pulse-designing protocol is more straightforward as the qubit instantaneous state analytically depends on the pulse parameters, and provides more degrees of freedom to construct and tailor the pulses envelope so as to achieve the required frequency selectivity, thus obtain high-fidelity manipulation between two arbitrary qubit superposition states. Pulses developed in counter-diabatic protocol can also achieve an arbitrary superposition state \cite{Vitanov1999} and can also be optimized along with unitary transformation \cite{Ban2018}, but involving some complexities. The state evolution depends strongly on adiabatic reference of pre-guessed pulses \cite{Alexandre2016,Ban2018}. Therefore, different systems with their own properties require unique shortcut design, which means that it is more obvious to choose inverse engineering for the required frequency selectivity, which is rather important for quantum control on qubits closely spaced in frequency.

The protocol proposed here can be in principle extended to a more generic n-level (n>3) system. For instance, the LR invariant of a four-level system has been found \cite{Uktan2012}, and shortcut to adiabaticity have been carried out by using 4D rotation \cite{Yichao2018, Yichao2018b}. Of course,  the similar protocol can be used and further optimized with respect to errors or noises, when the state evolution is parameterized.

Finally, we shall emphasize that the method presented here are applied to operations that start from a known initial state. This is different from the CHS pulses in dark state operations, which work for arbitrary operations on an unknown state. Optimally shortcut to adiabatic pulses for general gate operations in a three-level system need to be investigated. However, by dropping the requirement that they work on arbitrary inputs, fast and robust pulses with higher fidelity can be developed, which is helpful to initialize ancilla qubits used in error correction towards fault tolerant quantum computing. It can be applied to any quantum computing systems as well, where qubit is addressed in frequency.

\section*{Funding}

National Natural Science Foundation of China (NSFC) (61505133, 61674112); Natural Science Foundation of JiangSu Province (BK20150308); Key Projects of Natural Science Research in JiangSu Universities (16KJA140001); The International Cooperation and Exchange of the National Natural Science Foundation of China NSFC-STINT (61811530020); The project of the Priority Academic Program Development (PAPD) of Jiangsu Higher Education Institutions; A JiangSu Province Professorship; A Six Talent Peaks Project. S. K. acknowledges the support from the Swedish Research Council; the Knut and Alice Wallenberg Foundation; European Union`s Horizon 2020 research and innovation program (712721); NanOQTech and the Lund Laser Centre (LLC) through a project grant under the Lund Linneaus environment. This project has received funding from the European Union's Horizon 2020 research and innovation programme under grant agreement no 654148 Laserlab-Europe. X. C. acknowledges the support from National Natural Science Foundation of China (NSFC) (11474193); Shuguang Program (14SG35); STCSM (18010500400 and 18ZR1415500), and the Program for Eastern Scholar.

\newpage
\section*{APPENDIX}

\subsection*{A. Optimization of the pulses parameters $a_n$}

Parameters $a_n$ in all pulses shown above are optimized by checking the dependence of fidelity and off-resonant excitations in the $\left|\rm{0}\right>$ state on frequency detuning ($\Delta$) through manually scanning the $a_n$ parameters in three steps.

Step (1): Let $a_6 = a_8 = 0$, scan $a_2$. The 2-dimensional dependence of fidelity and off-resonant excitations on both $a_2$ and $\Delta$ is shown in Fig. 8(a) and Fig. 8(d), respectively. Ideally the fidelity within $\mid\Delta\mid \leq 170 \rm kHz$ range should be as close to 1 as possible, and the off-resonant excitations outside $\pm$3.5 MHz should be as low as possible. Fig. 8 (a) tells that $a_2 = -1.10$ is a good starting point.

Step (2): Let $a_2 = -1.10$ and keep $a_8 = 0$, scan $a_6$. Fig. 8(b) and Fig. 8(e) shows the results. $a_6 = [0.04, 0.12]$ is better, and we tentatively set $a_6 = 0.06$ and continue scan $a_8$ as follows.

Step (3): Let $a_2 = -1.10$ and $a_6 = 0.06$, scan $a_8$. Fig. 8(c) and 8(f) tell that $a_8 = [-0.06, 0.06]$ is better.

Repeat Step (3) for every value of $a_6$ in the range of [0.04, 0.12] in step of 0.02. Results show that parameters of $a_2 = -1.10$,  $a_6 = 0.06$, and $a_8 = 0.02$ provide the best performance. Average fidelity is 99.8\% over the detuning range of $\pm$340 kHz, and the off-resonant excitation is kept below 2.0\%.

To illustrate how the fidelity and off-resonant excitation change in response to above scanning steps, the dependence of them on $\Delta$ with optimized $a_n$ values get in each step are plotted in Fig. 9(a) and Fig. 9(b), respectively. The dash-dotted-blue line in each figure represents the case in step 1 where $a_2 = -1.10, a_6 = a_8 = 0$, the dashed-green line in Step (2) where $a_2 = -1.10, a_6 = 0.06$, and $a_8 = 0$, and the solid-red line in Step (3) where $a_2 = -1.10, a_6 = 0.06$ and $a_8 = 0.02$.

The optimization method used here is more intuitive to see the dependence of performance on the change in each parameters, but is less efficient. A faster optimization method using advanced optimization algorithms might be used and give different optimized parameters. However, the main results built under the theoretical frame presented in this work will not be altered by different parameters or optimization methods.

\newpage
\begin{figure}[htbp]
\begin{minipage}{8.5cm}
\includegraphics[width=6cm,height=4.5cm]{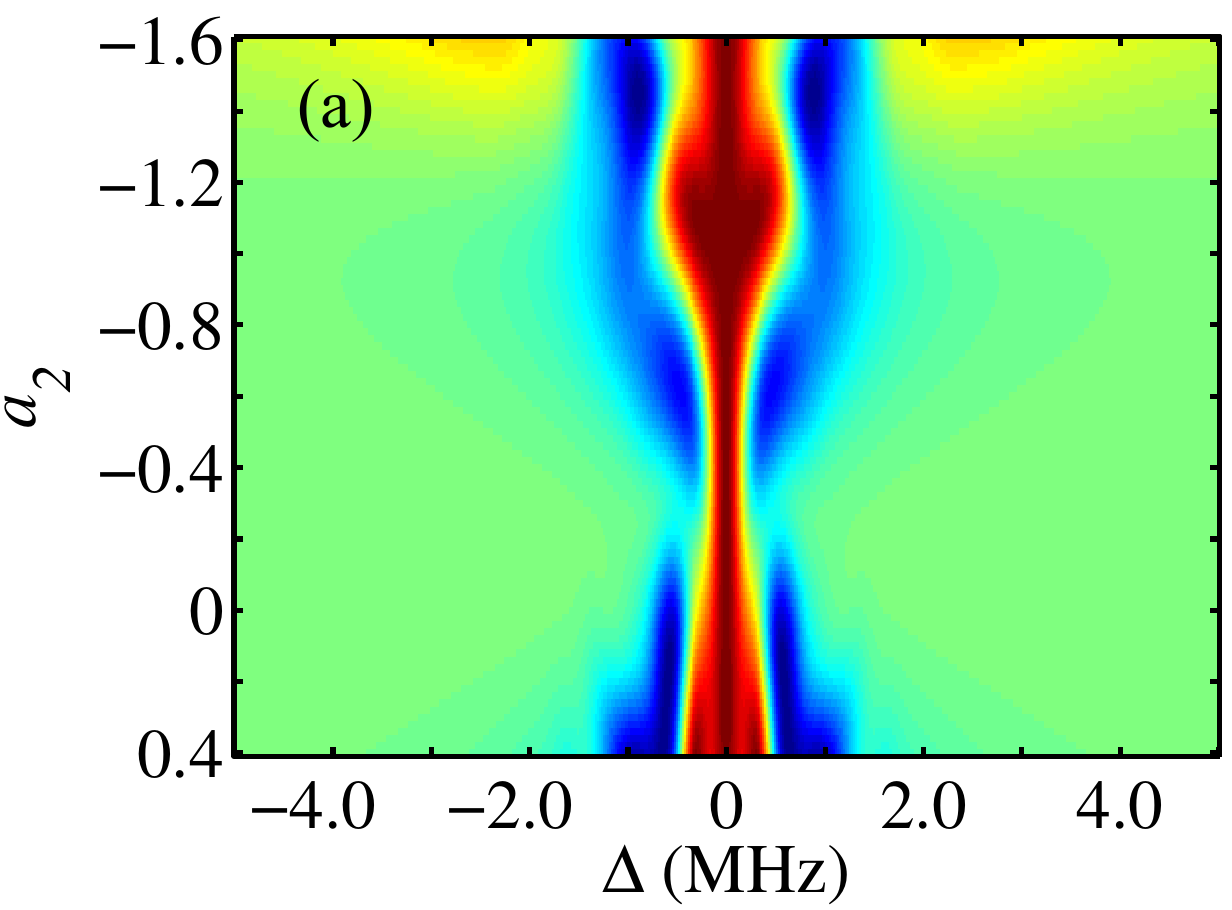}
\end{minipage}
\hfill
\begin{minipage}{8.5cm}
\flushleft
\includegraphics[width=6cm,height=4.5cm]{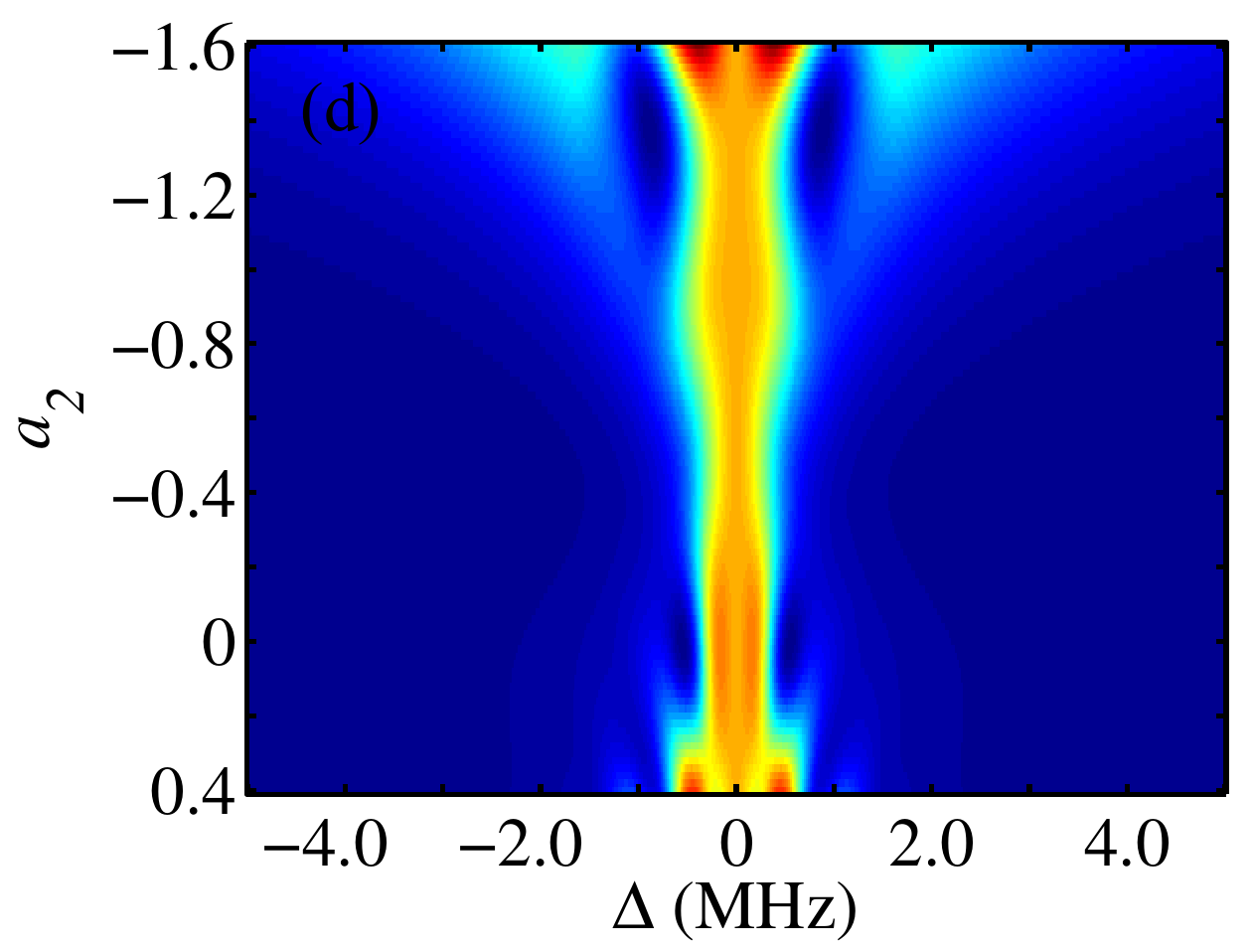}
\end{minipage}
\vfill  
\begin{minipage}{8.5cm}
\includegraphics[width=6cm,height=4.5cm]{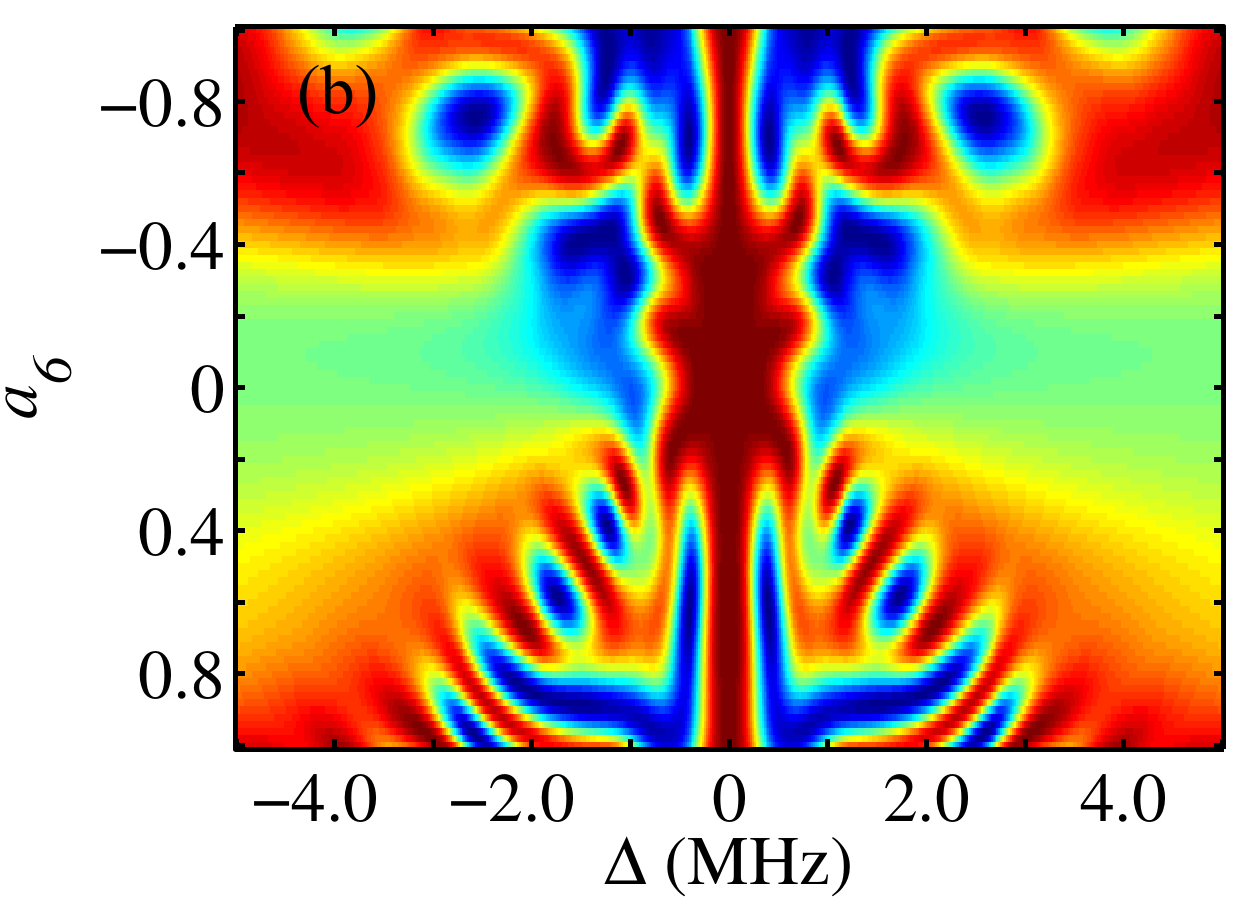}
\end{minipage}
\hfill  
\begin{minipage}{8.5cm}
\flushleft
\includegraphics[width=6cm,height=4.5cm]{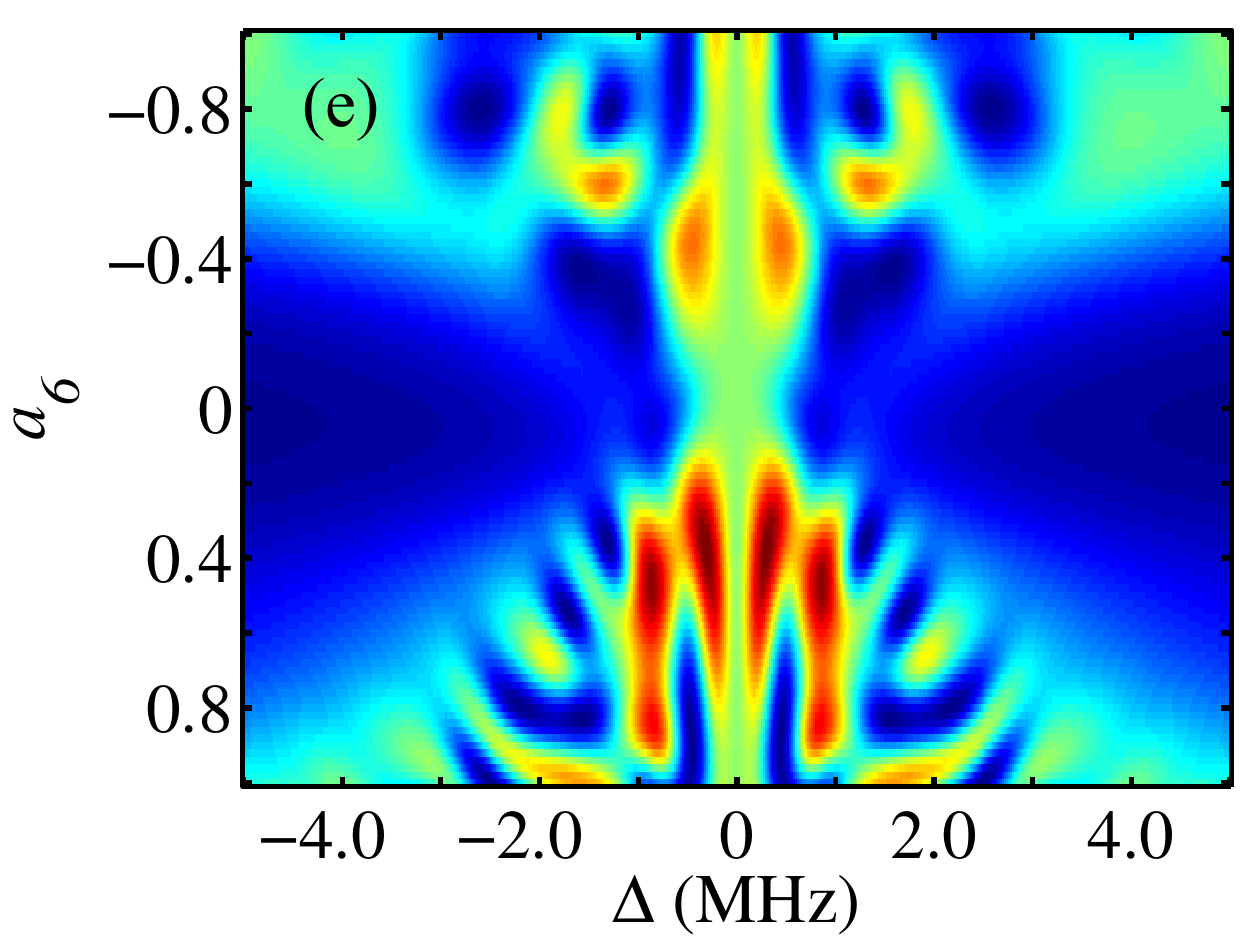}
\end{minipage}
\vfill
\begin{minipage}{8.5cm}
\includegraphics[width=7cm,height=4.5cm]{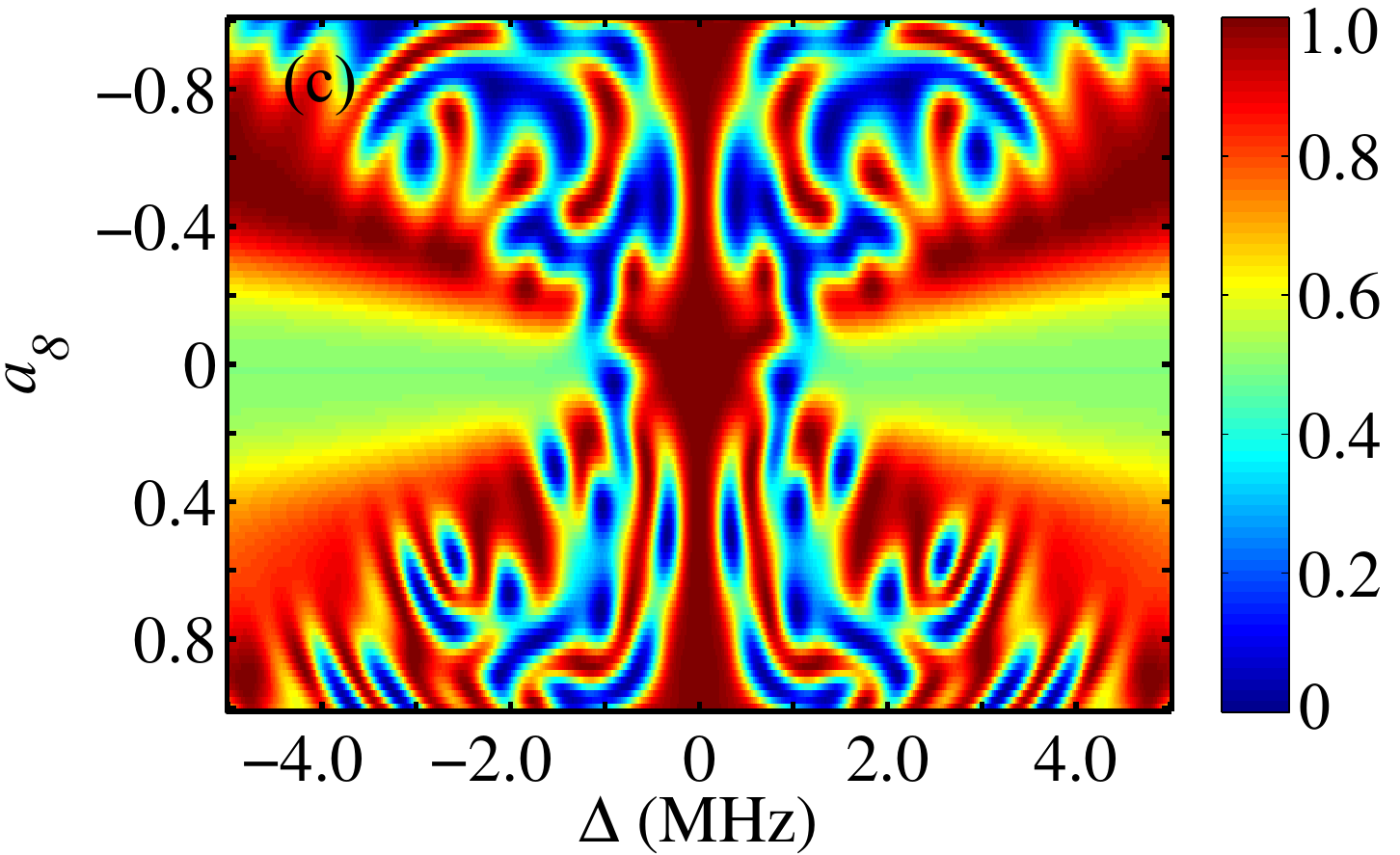}
\end{minipage}
\hfill
\begin{minipage}{8.5cm}
\flushleft
\includegraphics[width=7cm,height=4.5cm]{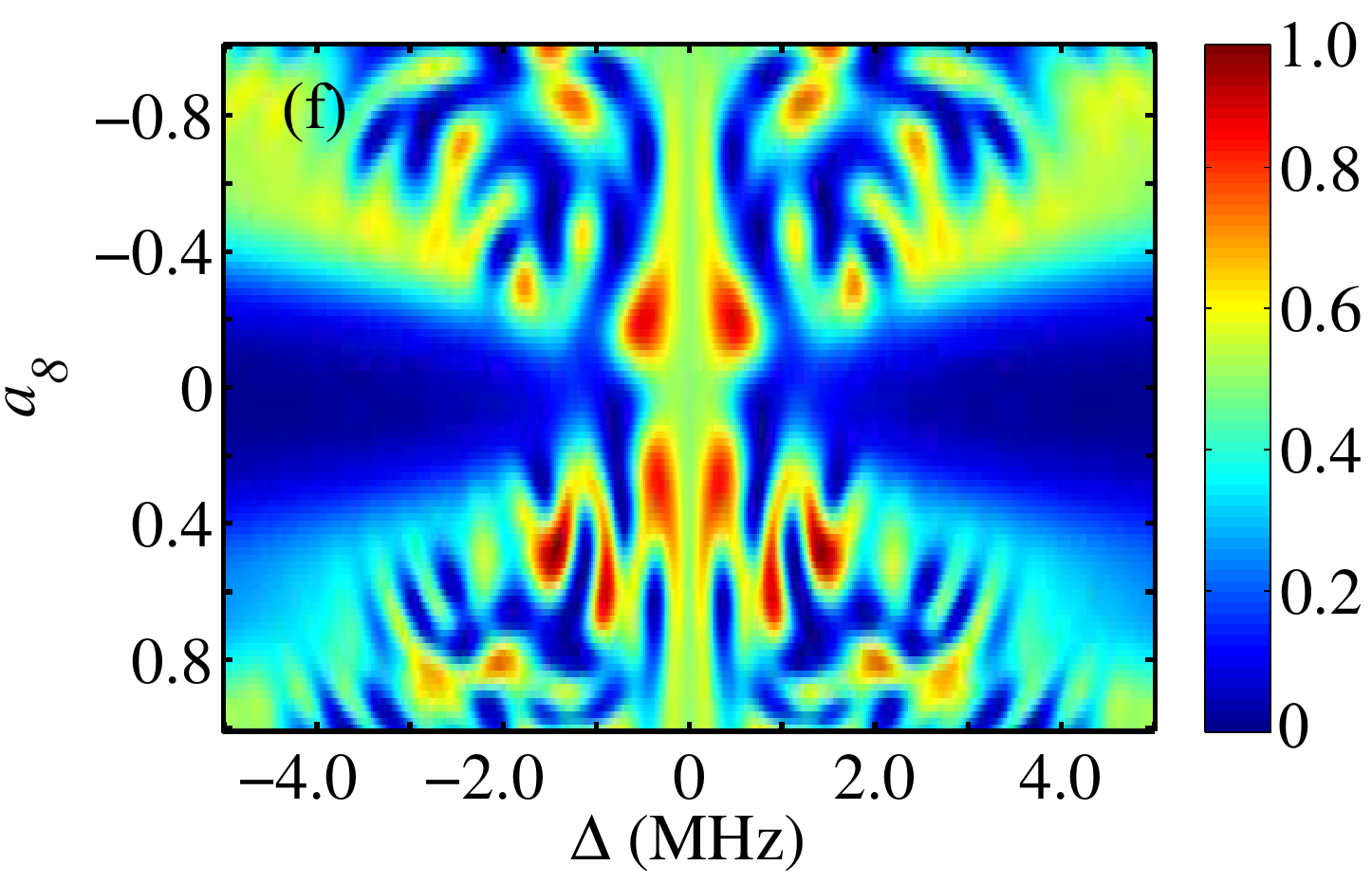}
\end{minipage}
	\caption{Optimization of $a_n$ values. Fidelity (a-c) and off-resonant excitations in $\left|\rm{0}\right>$ state (d-f) as function of frequency detuning while scanning pulse parameters. (a) and (d) scan $a_2$, $a_6 = a_8 = 0$; (b) and (e) Scan $a_6$, $a_2 = -1.10$, and $a_8 = 0$; (c) and (f) Scan $a_8$, $a_2$ = -1.10, $a_6$ = 0.06.}
	\label{fig.8}
\end{figure} 

\begin{figure}[htbp]
\begin{minipage}{8cm}
\centering
	\includegraphics[width=5.8cm,height=4cm]{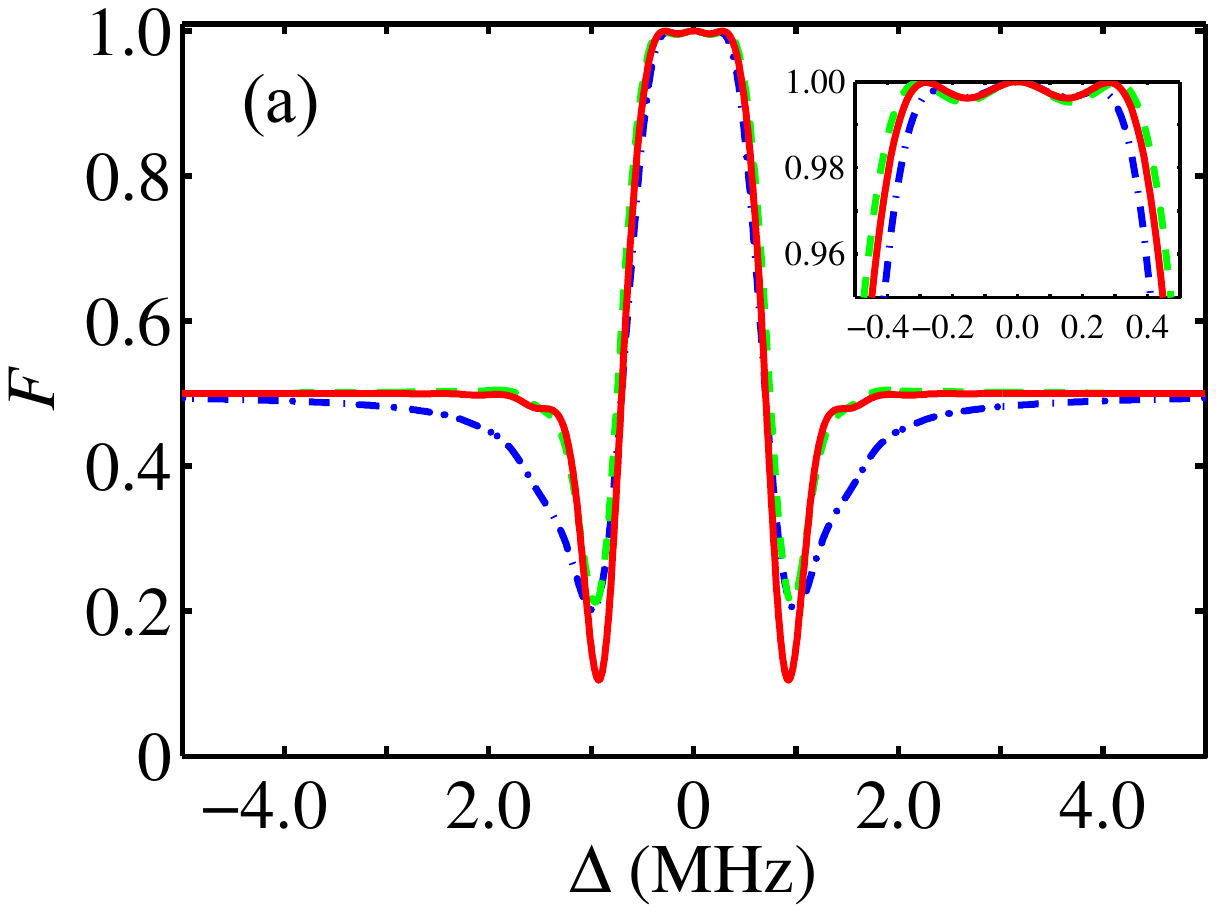}
\end{minipage}
\hfill
\begin{minipage}{8cm}
\centering
\includegraphics[width=5.8cm,height=4.2cm]{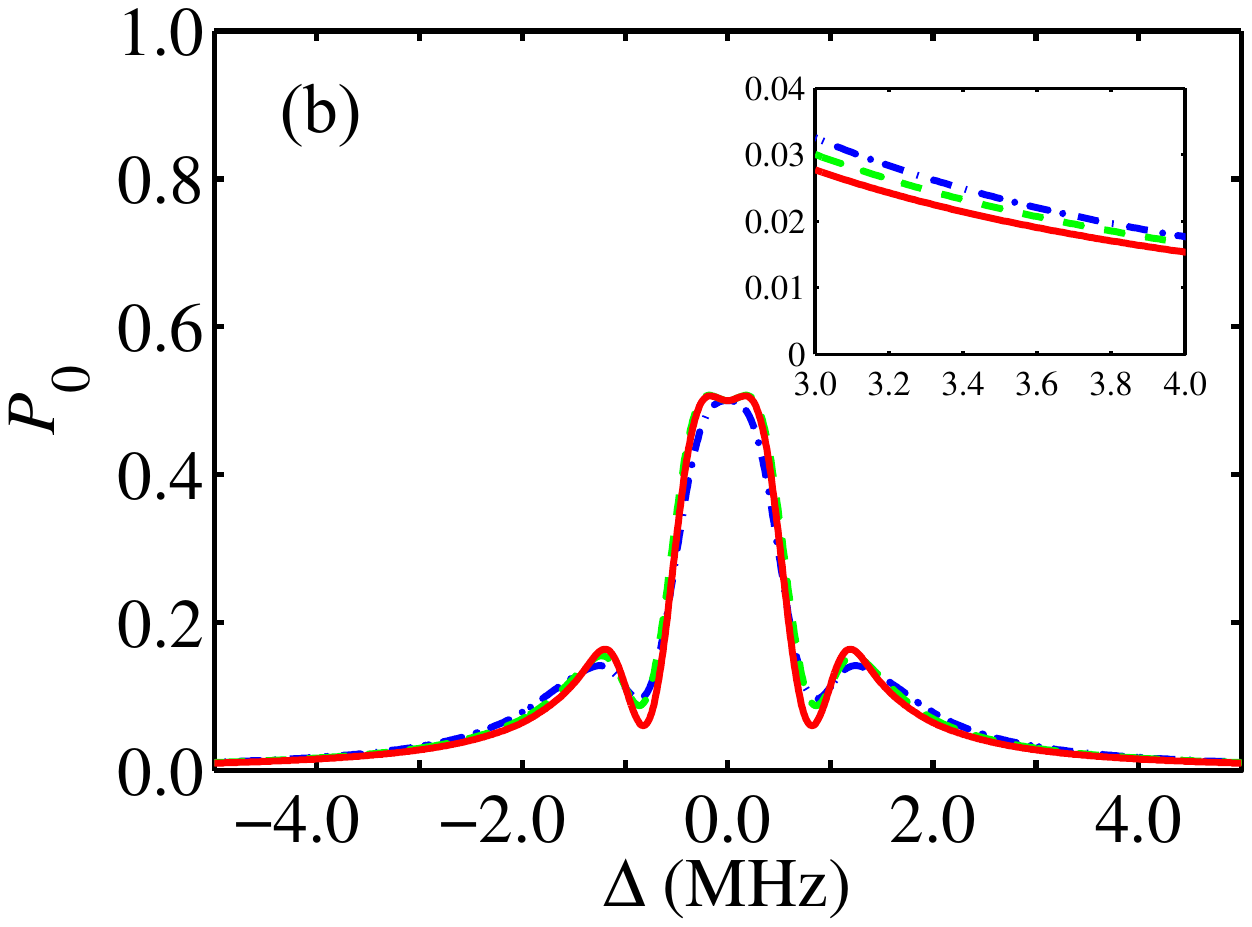}
\end{minipage}
\caption{Fidelity (a) and off-resonant excitations (b) on $\left|\rm{0}\right>$ state as a function of frequency detuning in the three scanning steps. Dash-dotted-blue lines: $a_2 = -1.1, a_6 = a_8 =0$. Dashed-green lines: $a_2 = -1.1, a_6 = 0.06$, and $a_8 = 0$. Solid-red lines: $a_2 = -1.1, a_6 = 0.06$, and $a_8 = 0.02$.}
\label{fig.9}
\end{figure} 

\newpage
\subsection*{B. Summary of all pulses for various operation tasks}

The pulses and optimized parameters for various operation tasks mentioned in the article are summarized in Table 1 below.

\begin{table} [htbp] 
  \centering
    \caption{The pulses and optimized $a_n$ for various operation tasks, where $\left|\psi_{\rm{in}}\right>$ ($\left|\psi_{\rm{tg}}\right>$) denotes the initial (target) state, $\left|\psi^{'}_{\rm{tg}}\right> = \cos\theta_a \left|1\right> + \sin\theta_a e^{i\varphi_a} \left|0\right>$, $\left|\psi^{'}_{\rm{in}}\right> = \cos\theta_b \left|1\right> + \sin\theta_b e^{i\varphi_b} \left|0\right>$, and $a_{1,3,5,7,} = 0$ in all cases.}\label{Table 1}
  \scalebox{0.95}{
  \begin{tabular}{c|c|c|l|l|l|c} 
  \hline
  \hline
  $\sharp$ & $\left|\psi_{\rm{in}}\right> $ & $\left|\psi_{\rm{tg}}\right> $ & ansatz & Rabi frequencies & optimized $a_n$ & system\\
  \hline
  \multirow{3}{*}{1} & \multirow{3}{*} {$\left|1\right>$} & \multirow{3}{*} {$\left|\psi^{'}_{\rm{tg}}\right>$} & $\gamma(t) = \frac{\pi}{t_f}\cdot t + \sum_{n=1}^{8} a_n \cdot \sin(\frac{n\pi}{t_f}\cdot t)$ & $\Omega_p = \dot{\gamma}(t)\cdot[(\pi-\theta_a)\cdot\cos{\gamma(t)}\sin{\beta(t)}+2\cos{\beta(t)}]$ & $a_2 = -1.1$ & \multirow{3}{*} {3-level} \\
   & & & $\beta(t) = \frac{\pi-\theta_a}{2}\cdot [1-\cos\gamma(t)]$ &$\Omega_s = \dot{\gamma}(t)\cdot[(\pi-\theta_a)\cdot\cos{\gamma(t)}\cos{\beta(t)}-2\sin{\beta(t)}]$ & $a_4 = 0.17$ & \\
   & & & $a_2 + 2a_4 + 3a_6 + 4a_8 = -0.5 $ & & $a_6 = 0.06$ & \\
   & & & & & $a_8 = 0.02$ & \\

  \hline
    \multirow{3}{*}{2} & \multirow{3}{*} {$\left|1\right>$} & \multirow{3}{*} {$i\left|e\right>$} & $\gamma(t) = -\frac{\pi}{2t_f}\cdot t + \sum_{n=1}^{8} a_n \cdot \sin(\frac{n\pi}{t_f}\cdot t)$ & $\Omega_p = 2\dot{\gamma}(t) $ & $a_2 = 0.5$  & \multirow{3}{*} {2-level}\\
   & & & $\beta(t)=0$ & $\Omega_s = 0$ & $a_4 = -0.335$ & \\
   & & & $a_2 + 2a_4 + 3a_6 + 4a_8 = 0.25 $ & &  $a_6 = 0.14$ & \\
   & & & & & $a_8 = 0$ & \\
   \hline
   \multirow{3}{*}{3} & \multirow{3}{*} {$\left|\psi^{'}_{\rm{in}}\right> $} & \multirow{3}{*} {$\left|1\right>$} & $\gamma(t) = -\frac{\pi}{t_f}\cdot t + \pi + \sum_{n=1}^{8} a_n \cdot \sin(\frac{n\pi}{t_f}\cdot t)$ & $\Omega_p = \dot{\gamma}(t)\cdot[(\pi-\theta_b)\cdot\cos{\gamma(t)}\sin{\beta(t)}+2\cos{\beta(t)}]$ & $a_2 = 1.06$ & \multirow{3}{*} {3-level} \\
   & & & $\beta(t)=\frac{\pi-\theta_b}{2}\cdot [1-\cos\gamma(t)]$ &$\Omega_s = \dot{\gamma}(t)\cdot[(\pi-\theta_b)\cdot\cos{\gamma(t)}\cos{\beta(t)}-2\sin{\beta(t)}]$ & $a_4 = -0.52$ & \\
   & & & $a_2 + 2a_4 + 3a_6 + 4a_8 = 0.5 $ & &  $a_6 = 0.16$ & \\
   & & & & & $a_8 = 0$ & \\
   \hline
   \hline
  \end{tabular}}
\end{table}
\newpage
\bibliographystyle{unsrt}  

\end{document}